\documentclass[fleqn]{aa}

\newif\ifpdf
\ifx\pdfoutput\undefined
   \pdffalse              
\else
   \pdfoutput=1           
   \pdftrue
\fi
\ifpdf
  \usepackage[pdftex]{graphicx}
  \usepackage[bookmarks=true]{hyperref}
  \hypersetup{%
        bookmarksopen=true,
        colorlinks=true,
        pdfpagemode=UseOutlines,
        pdftitle={Title},
        pdfsubject={Subject},
        pdfauthor={Dr.~C.~Zier}
  }
  \usepackage{thumbpdf}
\else
  \usepackage[dvips]{graphicx}
\fi

\usepackage{latexsym}
\usepackage{amssymb}
\usepackage{amsmath}
\usepackage{natbib}
\usepackage{journals}
\usepackage{graphicx}
\newcommand{\mli}[2]{#1 _{\rm{#2}}}
\newcommand{\VEC}[1]{\mbox{\boldmath${\mathrm{#1}}$\unboldmath}}
\newcommand{\mean}[1]{\langle #1 \rangle}
\newcommand{\ut}[1]{\,{\rm #1}}

\newcommand{\eqn}[1]{Eq.~(\ref{#1})}
\newcommand{\fgr}[1]{Fig.~\ref{#1}}
\def\BHBidx{{\bullet\!\bullet}}
\def\P1{Paper\,{\footnotesize I}}
\def\p1{paper\,{\footnotesize I}}

\def\d{{\rm d}}
\def\picwidth{88mm}
\begin{document}

   \title{Binary black holes and tori in AGN}

   \subtitle{II. Can stellar winds constitute a dusty torus?}

   \author{C. Zier
          \and
          P. L. Biermann 
          }

   \offprints{C. Zier}

   \institute{Max-Planck Institut f\"ur Radiostronomie (MPIfR), Bonn,
              Auf dem H\"ugel 69, D-53121 Bonn\\
              email: chzier@mpifr-bonn.mpg.de
             }

   \date{\today}

   \abstract{In this second paper, in a series of two, we determine
     the properties of the stellar torus that we showed in the first
     paper to result as a product of two merging black holes. If the
     surrounding stellar cluster is as massive as the binary black
     hole, the torque acting on the stars ejects a fraction which
     extracts the binary's angular momentum. After the black holes
     coalesced on scales of $\sim 10^7\ut{yr}$, a geometrically thick
     torus remained. In the present article we show that a certain
     fraction of the stars has winds, shaped into elongated tails by
     the central radiation pressure, which are optically thick for
     line of sights aligned with them. These stars are sufficiently
     numerous to achieve a covering factor of $1$, so that the
     complete torus is optically thick. This patchy structured torus
     is then compared with observations. We find the parameters of
     such a torus to be just in the right range in order to explain
     the observed large column densities in AGN and their temporal
     variations on time scales of about a decade. Within this model
     the broad absorption line quasars can be interpreted as quasars
     seen at intermediate inclination angles, with the line of sight
     grazing the edge of the torus. The half-opening angle of the
     torus is wider for major mergers and thus correlates with the
     central luminosity, as has been suggested previously. In this
     picture the spin of the merged black hole is possibly dominated
     by the orbital angular momentum of the binary. Thus the spin of
     the merged black hole points into a new direction, and
     consequently the jet experiences a spin-flip according to the
     spin-paradigm. This re-orientation could be an explanation for
     the X-shaped radio galaxies, and the advancing of a new jet
     through the ambient medium for Compact Symmetric Objects.

}

   \maketitle

%

\section{Introduction}
\label{s_intro}

Active galactic nuclei (AGN) produce very high luminosities in a very
restricted volume, probably via an accretion disk around a massive
black hole (BH) in which shear stresses cause matter to sink down in
the potential of the BH. These luminous nuclei appear in many
different flavours what the unification scheme traces back to a
nonspherical symmetry of the nucleus, which is spatially not resolved.
Thus the same object looks different if viewed from different
directions, leading to various classifications.

\begin{figure*}[t]
\vspace{-8mm}
\ifpdf
  \resizebox{182mm}{!}{\includegraphics[angle=-90]{t_agn_scetch.pdf}}%
\else
  \resizebox{182mm}{!}{\includegraphics[angle=-90]{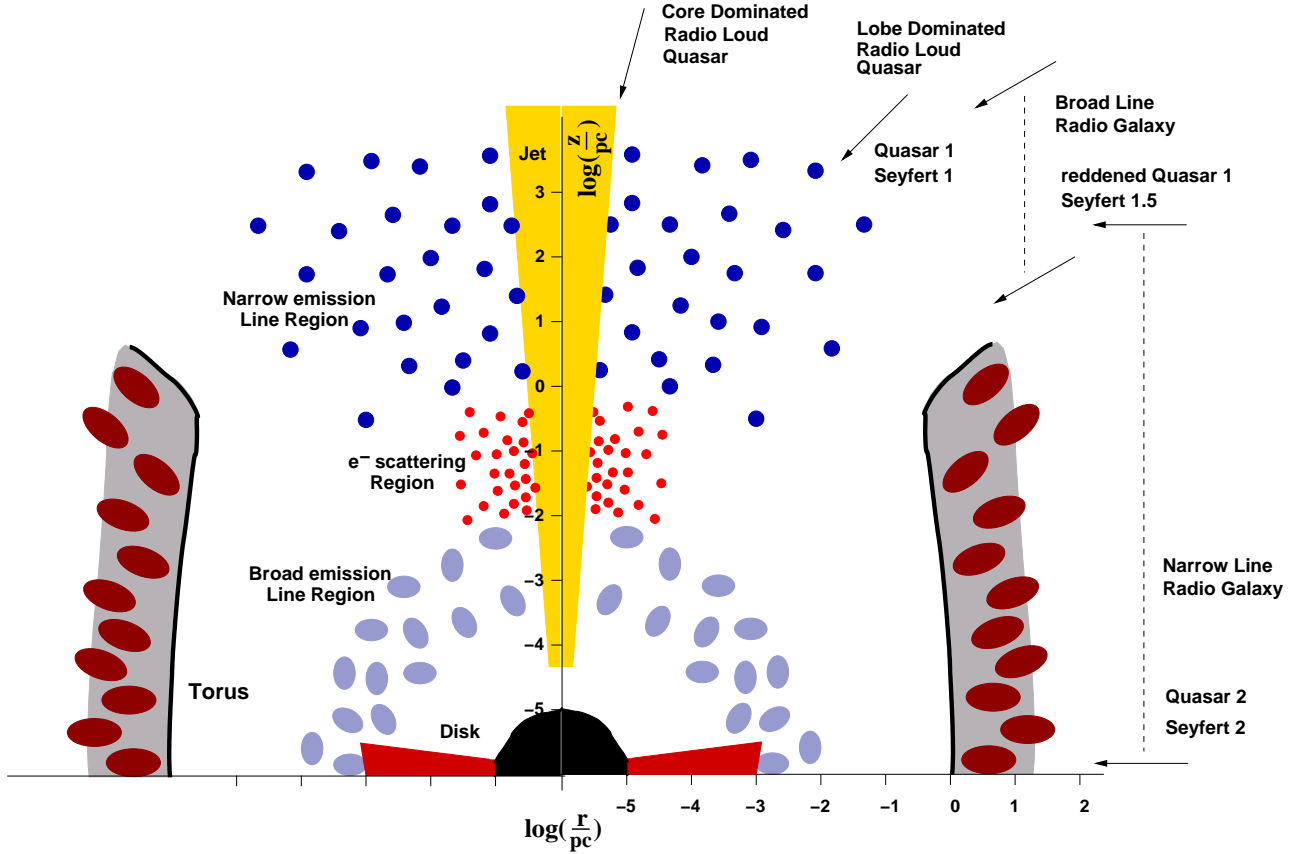}}%
\fi
\vspace{-8mm}
\caption[]{ A sketch of the cylindrically symmetric AGN according to
  the description in the introduction and to the picture of a patchy
  torus in AGN, developed in \p1 and this work, is shown. The cut
  shows the $r-z$-plane, both axes logarithmically scaled to $1\,{\rm
    pc}$. The basic constituents are the central BH with a surrounding
  accretion disk, the jet perpendicular to the disk and the torus
  encircling this configuration. The dark patches in the torus
  indicate the clouds, made up by stellar winds in our proposed
  concept. The locations of the BLR, ESR and NLR are shown on the
  left, as well as the appearance of the nucleus as a function of the
  angle of the LOS to the jet-axis (on the right). See text for
  details. Note that a strictly spherical distribution is almost a
  square box in this diagram.}
\label{f_agn_sketch}
\end{figure*}

Such an AGN we tried to illustrate in Fig.~\ref{f_agn_sketch} in
double logarithmic scales. In the very center is a massive BH of about
$10^8\,M_\odot$ with a Schwarzschild radius of about $10^{-5}\ut{pc}$.
The observed rapid variation of \mbox{X-rays} on scales of $10^4\,{\rm
  s}$ restricts their origin to a volume of about $3\times
10^{12}\ut{m}$ radial extension, which corresponds to the last stable
orbit of a black hole of $\sim 3\times 10^8\,M_\odot$.  The
\mbox{X-rays} are thought to be generated near the inner regions of
the accretion disk, which produces the optical-UV emission in the
outer parts, extending to about $10^{-3}\ut{pc}$.  On scales up to
$0.01\ut{pc}$ the broad emission line region (BLR) is found
\citep{peterson93} with line widths corresponding to
doppler-broadening with velocities of $1500$ up to $30\,000\ut{km/s}$,
probably caused by line emission in clouds moving at these speeds.
While these lines are varying on scales of weeks or months, in
agreement with the extension of the BLR, the narrow emission lines do
not show such variations.  Their line widths imply velocities of the
order of $1000\ut{km/s}$ and hence place the narrow emission line
region (NLR) at distances up to the kilo parsec scale.
  
In the infrared (IR) the spectrum shows a bump between $2\ut{\mu m}$
and $1\ut{mm}$ which is thought to be due to re-radiation by dust in a
distance of some parsec from the center \citep{chini89,sanders89b} and
which is distributed in a toroidal volume around the nucleus, whose
symmetry axis is aligned with that of the accretion disk (see
Fig.~\ref{f_agn_sketch}).
  
The observed AGN can basically be divided in two different types, with
Type~1 objects exhibiting all the features described above. Therefore
in these objects the central parts are directly seen without absorbing
matter in the line of sight (LOS). On the other hand in sources of
Type~2, there is no convincing evidence for the directly visible
optical-UV emission and the soft X-rays seem to be heavily absorbed
\citep{lawrence82,mushotzky82}. No broad emission lines are detected.
The unification scheme relates the different types to different
orientations rather than intrinsic differences: The Type~2 classified
nuclei are seen at large inclination angles and consequently the LOS
is blocked by the torus so that only the NLR is freely visible and in
fact indistinguishable from that in Type~1 objects.
  
Evidence for such obscuring tori is deduced from observed sharply
defined conical/biconical nebulae which are suggestive of shadowing
(\citet{antonucci93} and references therein).  The deficit of observed
ionizing photons relative to what is needed to ionize the nebulae
\citep{kinney91} evidently supports the idea of matter obscuring the
central source, which is not obscured at the aspect angles where the
ionized nebulae are located. Hence the ionizing photons escape from
the center in a cone so that the optical, UV and probably soft X-ray
radiation is collimated by the torus, escaping anisotropically through
the polar cap regions.
  
Even stronger support comes from spectrapolarimetry, which reveals in
the polarized spectrum of Type~2 sources a clear spectrum of Type~1
\citep{antonucci85,miller91}. The power-law continuum has the same
degree of polarization as the permitted broad lines, which also have
normal equivalent width. The degree of polarization is found to be
almost wavelength independent. This is interpreted as the reflection
of the obscured Type~1 continuum by free electrons in the opening of
the torus (\citet{antonucci93} and references therein).  This electron
scattering region (ESR) is located somewhere near to the BLR and
extends to scales of the height of the torus \citep{taniguchi99a}, see
Fig.~\ref{f_agn_sketch}.

Thus for a small viewing angle with the source being seen almost face
on all features of a Type~1 spectrum are visible. As the viewing angle
increases the LOS finally grazes the edge of the torus and the
spectrum becomes reddened. The central regions gradually become more
obscured by the torus till the broad lines disappear and the source is
classified as Type~2 when it is seen near edge on.
  
As indicated in Fig.~\ref{f_agn_sketch}, there are more criteria than
the orientation to distinguish between the different objects, such as
the luminosity or the radio properties, with radio loud galaxies
having a strong jet streaming along the symmetry axis.  For a much
more detailed explanation of the unified scheme see for example the
reviews by \citet{antonucci93}, \citet{urry95} and \citet{wills99}.

The torus is a very important component in this scheme.  To support
its vertical thickness previous models either use magnetic pressure
\citep{lovelace98} or radiation pressure \citep{pier92b}, which faces
some problems for low luminosity central sources and smoothly
distributed dust. Here we want to further develop our idea, introduced
in a previous paper \citep{zier01}, that this torus is comprised of
the winds of stars moving in the potential of a merging binary black
hole (BBH), allowing them to maintain the geometrical thickness of the
torus.

In that article we simulated the evolution of a stellar cluster under
the influence of a massive binary black hole in its center, based on
the two widely accepted assumptions that (a) galaxies with an active
nucleus harbour a supermassive black hole in their center and (b) that
galaxies frequently merge. The results clearly showed that the binary
merges on scales of some $10^7\ut{yr}$ due to ejection of a fraction
of stars, if the surrounding stellar cluster is about as massive as
the binary, and ensuing emission of gravitational radiation. The stars
which remained bound indeed are found to be distributed geometrically
in the volume of a thick torus. In this second paper we will
investigate, wether such a stellar torus can explain the presence of
the ubiquitous torus surrounding apparently all AGN.

The next section will briefly review the constraints put on the
properties of the torus by observations.


\section{Properties of the torus}
\label{s_torus}

Although the current telescopes are not able to spatially resolve the
torus in AGN, conclusions could be drawn from observations in various
wave bands about the torus' geometry as well as the matter it is made
of.

\subsection*{The inner radius}

Absorption of the primary optical-UV radiation by dust, which reemits
it in the IR, heats the dust to its evaporation temperature ($\sim
1500\ut{K}$). Inside the evaporation radius dust can not exist and
\citet{lawrence91} assumed it to be the inner radius of the dusty
torus. This radius is about $1\ut{pc}$, corresponding to the distance
where the BBH becomes hard \citep{milos01} and what also marks the
inner radius of the torus that we obtained in our simulations (\p1).
This value is also in agreement with the inner radius obtained by
\citet{krolik88} from the balance between cloud evaporation by central
radiation and inflow by dissipative processes in the torus. In several
Type~1 AGN variations in the near-infrared (NIR) emission have been
observed to lag those at shorter wavelengths
\citep{clavel89,sitko91,baribaud92}.  These time delays place the IR
emission source in a distance from the central engine which is
consistent with that of optically thin nuclear heated dust at its
evaporation temperature.

To explain the bump in the near-infrared in terms of thermal dust
emission, a range in the grain temperature is required
($1500\ut{K}\lesssim T_{\rm dust}\lesssim 30\ut{K}$), since isothermal
dust close to its evaporation temperature yields a narrower bump than
that observed \citep{sanders89b,haas00}. Thus the torus' temperature
has to decrease from the inner edge to its outer regions
\citep{niemeyer93}, but a very thick and compact torus
\citep{pier92,pier93} is not able to provide such a broad temperature
range. Instead a clumpy torus, as suggested by \citet{krolik88}, has
the advantage that dust in clouds can survive more easily the strong
radiation field, and such a structure of the torus tends to increase
the dust temperature in the outer parts \citep{efstathiou95} where the
clouds are exposed to central radiation through gaps in the inner
parts of the patchy torus. This seems also most promising to
\citet{haas00} in order to explain the spectral energy distribution
(SED).

\subsection*{Column density $N_{\rm H}$}
On the basis of X-ray observation lots of evidence has been
accumulated that the column density is in the range $10^{22-24}
\ut{cm}^{-2}$ (\citet{mulchaey92} for UV-detected Seyfert~2s), or
$1-8\times 10^{24}\ut{cm}^{-2}$ \citep{krolik88}. A more recent study
of the X-ray observations of a sample of Seyfert 2 galaxies in the
$0.1-100\ut{keV}$ spectral range with \emph{BeppoSAX} has been
performed by \citet{maiolino98}. The sources they studied were
selected according to their [OIII] optical emission line flux after
correction for the extinction deduced from the Balmer decrements.
Hence this sample is thought to be fairly isotropic and avoids a bias
against heavily obscured nuclei, as is probably the case in former
samples. \citet{maiolino98} find all the sources to be absorbed by
column densities larger than $4\times 10^{23}\ut{cm}^{-2}$ and most
even appear to be thick to Compton scattering with $N_{\rm H}\gtrsim
10^{25}\ut{cm}^{-2}$. Hence their results provide further and strong
evidence for the unification scheme and indicate that the obscuration
in Type~2 AGN is probably much higher than deduced from earlier X-ray
surveys.

Such high column densities are also deduced from radio data.
\citet{conway95} ascribe the absorption of the broad HI $21\ut{cm}$
line towards the compact radio nucleus of the FR II radio galaxy Cyg~A
to the obscuration by a torus in about $10\ut{pc}$ distance to a
central BH of $\sim 10^8\,M_\odot$. Assuming a spin temperature in the
range $8000-16\,000\ut{K}$ the authors obtain a column density in the
range $2-4\times 10^{23}\ut{cm}^{-2}$, in very good agreement with the
column $3.75\times 10^{23}\ut{cm}^{-2}$ deduced from previous X-ray
spectroscopy by \citet{ueno94}.

Both, X-ray and radio data show also evidence for a patchy structured
torus \citep{boller02,krichbaum98}, in agreement with the explanations
for the IR range of the previous section.

\subsection*{Opening angle}

The observed sharply defined conical/biconical nebulae in some AGN are
suggestive of shadowing (\citet{antonucci93} and references therein).
These ionization-cones, being located in the opening of the obscuring
torus and aligned with the inner radio jets \citep{nagar99}, represent
a lower limit of the half opening angle of torus in the range
$35^\circ$ to $50^\circ$
\citep{pogge89,wilson93,wilson94,macchetto94}.

Using the statistics of the observed number ratios of Seyfert 1 and 2s
\citep{osterbrock88,huchra92,barthel89,willott99} a half-opening angle
of the torus in the range $30^\circ - 60^\circ$ is obtained. This
matches very well the angles of the ionization-cones inside the
opening of the torus.

\subsection*{Covering factor}
According to the opening angle of the torus, it screens a considerable
fraction of the solid angle, seen from the central source, and is
geometrically thick. Blocking the view to the nucleus in Type~2 AGN,
the torus, if comprised of individual clouds, must be optically thick
for most of the lines of sight penetrating it. Thus the clouds
constituting the torus must achieve a covering factor of about one, if
a single cloud is optically thick, and more otherwise.

Hence the observations demand a clumpy-structured torus with an inner
radius of about $1\ut{pc}$, that is optically and geometrically thick
with a half-opening angle of $\sim 45^\circ$ and therefore has a
covering factor of order unity.


\section{Kinematics and density distribution of the stars}
\label{sc_kinematics}

\begin{figure}
\ifpdf
  \resizebox{\picwidth}{!}{\includegraphics{f_dx2_rho_q1.pdf}}%
\else
  \resizebox{\picwidth}{!}{\includegraphics{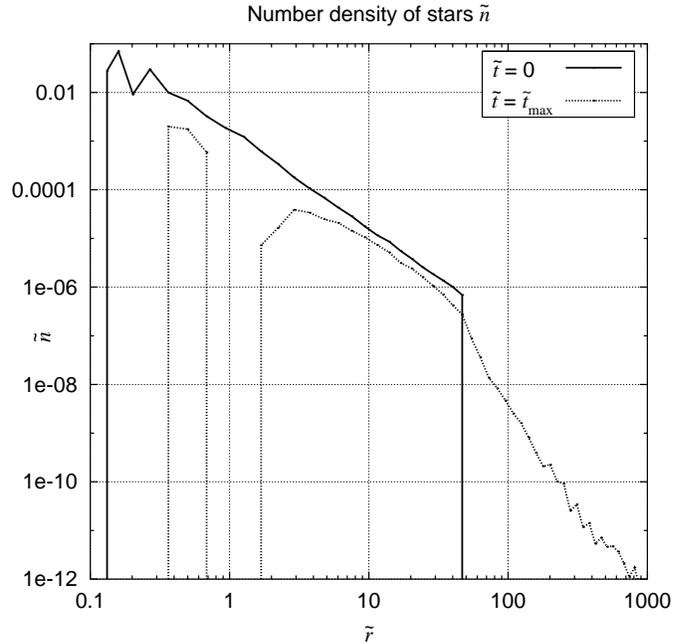}}%
\fi
\caption[]{
  The stellar number density distribution is shown at the beginning
  ($\tilde{t}=0$) and end ($\tilde{t}=\mli{\tilde{t}}{max}$) of the
  simulation. Finally, at $\mli{\tilde{t}}{max}$, the bound star
  distribution follows a powerlaw with index $-4$ in the heated region
  ($\tilde{r}\gtrsim 50$), while an index $\sim -2$ is maintained in
  the range $10\lesssim\tilde{r}\lesssim 50$. The inner parts are
  scoured out due to ejection of stars by the binary, and a maximum
  emerges at $\tilde{r}\approx 3$, showing the torus-like
  configuration the stars assume. For $\tilde{r}\lesssim 0.7$ a cusp
  of stars bound to $M_1$ only is left.  $\tilde{r}$ is normalized to
  $1\,{\rm pc}$ and $\tilde{n}$ such that the area underneath the
  solid line is $1$.  }
\label{f_dx2_rho_q1}
\end{figure}

Our simulation of a stellar distribution in the potential of the
binary (discussed in detail in \p1, \citet{zier01}) has been performed
for three different mass-ratios of the BHs ($q=1,\,10,\,100$), with
\[
q = \frac{M_1}{M_2}\,,
\]
where $M_1\ge M_2$, always. While we focused on the ejected fraction
of the stars in that article we will discuss in this section the bound
population. For clarity we first briefly summarize just a few results
from \p1, using a mass-ratio of the BHs of $q=1$.

\subsection{Dynamics of the bound stars}
\label{sc_bound-kin}

In order to write the equations in their dimensionless form, denoted
by a tilde ``$\tilde{\quad}$'' on top of the quantities, we used the
following normalization parameters:
\begin{eqnarray*}
r_0 & = & a = 1\ut{pc}\,,\\
t_0 & = & \sqrt{\frac{r_0^3}{G(M_1+M_2)}}\,,\\
L_0 & = & \frac{m a^2}{t_0}\,.
\end{eqnarray*}
The semi-major axis of the BBH is denoted by $a$, and for the mass of
a typical star $m$ we used one solar mass.

\begin{figure}
\ifpdf
  \resizebox{\picwidth}{!}{\includegraphics{f_dx2_beta_q1.pdf}}%
\else
  \resizebox{\picwidth}{!}{\includegraphics{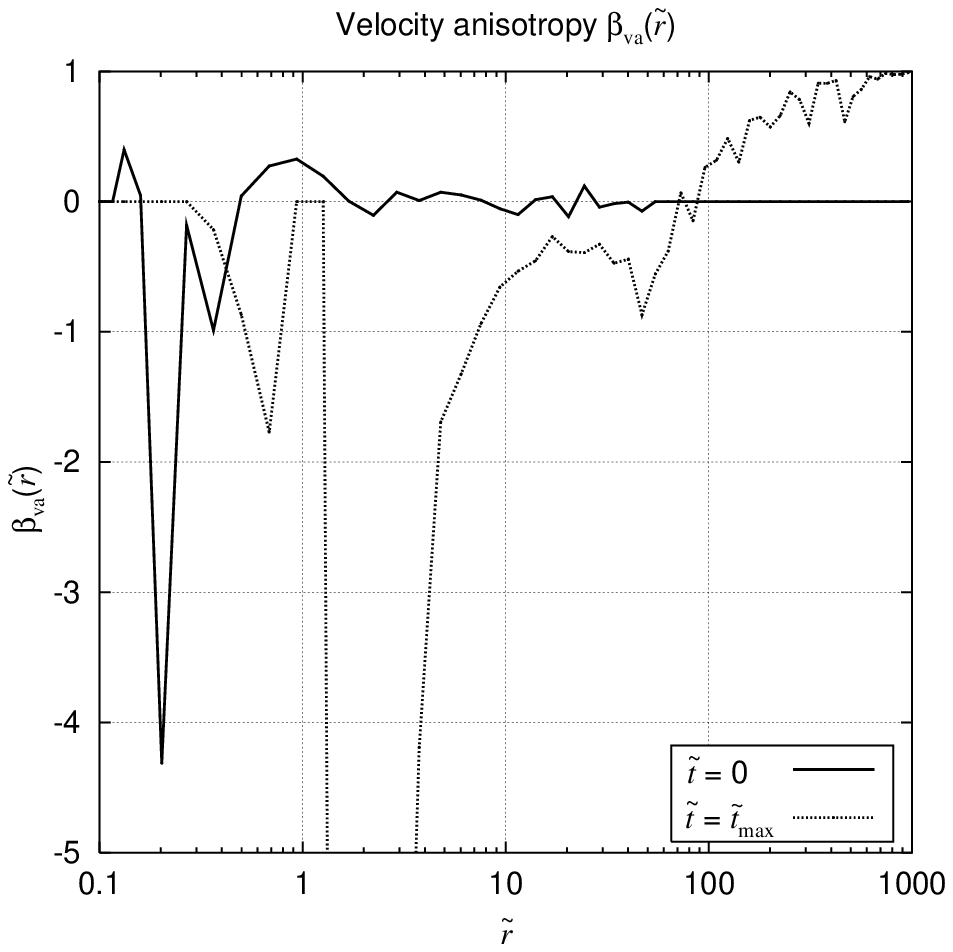}}%
\fi
\caption[]{
  The initial conditions ($\tilde{t}=0$) of the star cluster have been
  set so that the velocity anisotropy
  $\mli{\beta}{va}=1-\frac{\mean{\tilde{v}_\phi^2} -
    \mean{\tilde{v}_\theta^2}}{2\mean{\mli{\tilde{v}}{r}^2}}$ is
  neither radially ($\ge 1$) nor tangentially ($\le 1$) anisotropic.
  At $\tilde{t}=\mli{\tilde{t}}{max}$ the velocity anisotropy of the
  bound stars is a strongly increasing function of the radius. In the
  range $r\lesssim 50$ they are tangentially anisotropic, while in the
  heated region ($r\gtrsim 50$) the stars are moving on radially
  anisotropic orbits.}
\label{f_dx2_beta_q1}
\end{figure}

During the simulation the initial singular isothermal sphere evolves
finally into a distribution which is depleted in the inner regions
(see Fig.~\ref{f_dx2_rho_q1}). The stellar density peak at about
$\tilde{r}= 0.5$ is made up by a fraction less than $0.4\%$ of the
final distribution and is bound to the primary black hole only. Since
we expect this cusp to be much shallower if we would have taken into
account the shrinking of the binary in the calculations, we will not
address it further in our discussion.

After a gap in a distance corresponding to the radial distance of the
secondary black hole to the center, the density increases again and
peaks at $\tilde{r}\sim 3$. In the range $10\lesssim \tilde{r}\lesssim
50$ the density follows the initial powerlaw of the isothermal sphere
($\tilde{n}\propto \tilde{r}^{-2}$) and for radii bigger than the
outer limit of the initial distribution ($\tilde{r}\gtrsim 50$)
approximates a powerlaw with index $-4$.

\begin{figure}[t]
\ifpdf
  \resizebox{\picwidth}{!}{\includegraphics{f_dx2_ldotr_q1.pdf}}%
\else
  \resizebox{\picwidth}{!}{\includegraphics{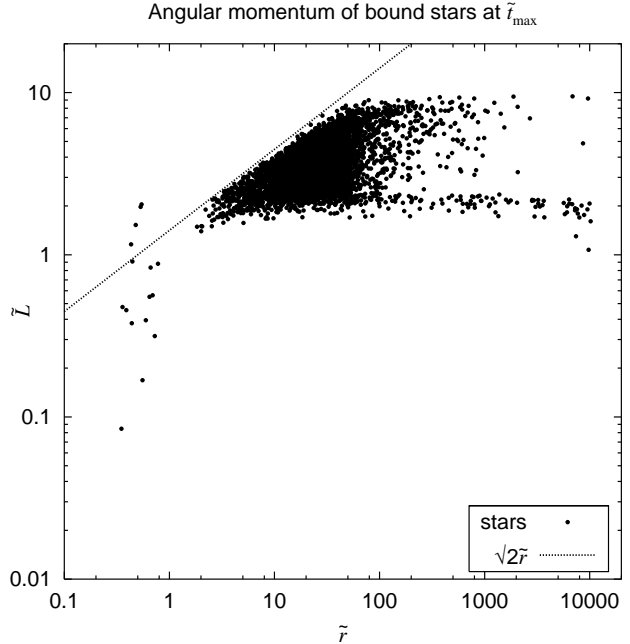}}%
\fi
\caption[]{Plotting the angular momentum $\tilde{L}$ versus the
  distance $\tilde{r}$ at $\mli{\tilde{t}}{max}$ shows all stars to
  respect the upper limit indicated by the dotted line. Only some
  stars in the cusp region ($\tilde{r}<1$) have larger angular
  momenta, because they are bound to one BH only, with the other
  perturbing their orbits, so that the assumption of a point-mass
  potential does not apply at these small distances. For
  $\tilde{r}\gtrsim 50$ two populations can be distinguished, one
  clustering around $\tilde{L}\approx 2$ (heated stars) and the other
  having an upper limit of $\tilde{L}\approx 10$, fading towards
  smaller values (relaxation process). See text for details.}
\label{f_dx2_ldotr_q1}
\end{figure}

In \p1 we showed that the stars with small angular momenta, i.e. small
pericenters, get close to the orbit of the secondary black hole where
they are strongly interacting with both BHs, so that they become
ejected.  Only stars moving on orbits with sufficiently large
pericenters ($\tilde{r}_{-}\gtrsim 2$) can remain in the central
regions. The smaller their radii, the more they are tangentially
anisotropic (see Fig.~\ref{f_dx2_beta_q1}) and the less eccentric are
their orbits.  A fraction of the stars which are interacting with the
BHs will be heated to larger distances and stays bound to the binary
instead of becoming ejected. Such stars did not gain as much energy as
the ejected ones and contribute to the distribution at distances
larger than the extension of the initial distribution, which
terminates at $\tilde{r}=50$ (see Fig.~\ref{f_dx2_rho_q1}).

\begin{figure*}[t]
\hspace{1mm}%
\ifpdf
  \resizebox{\picwidth}{!}{\includegraphics{f_gs_q1_t8_x_new.png}}%
\else
  \resizebox{\picwidth}{!}{\includegraphics{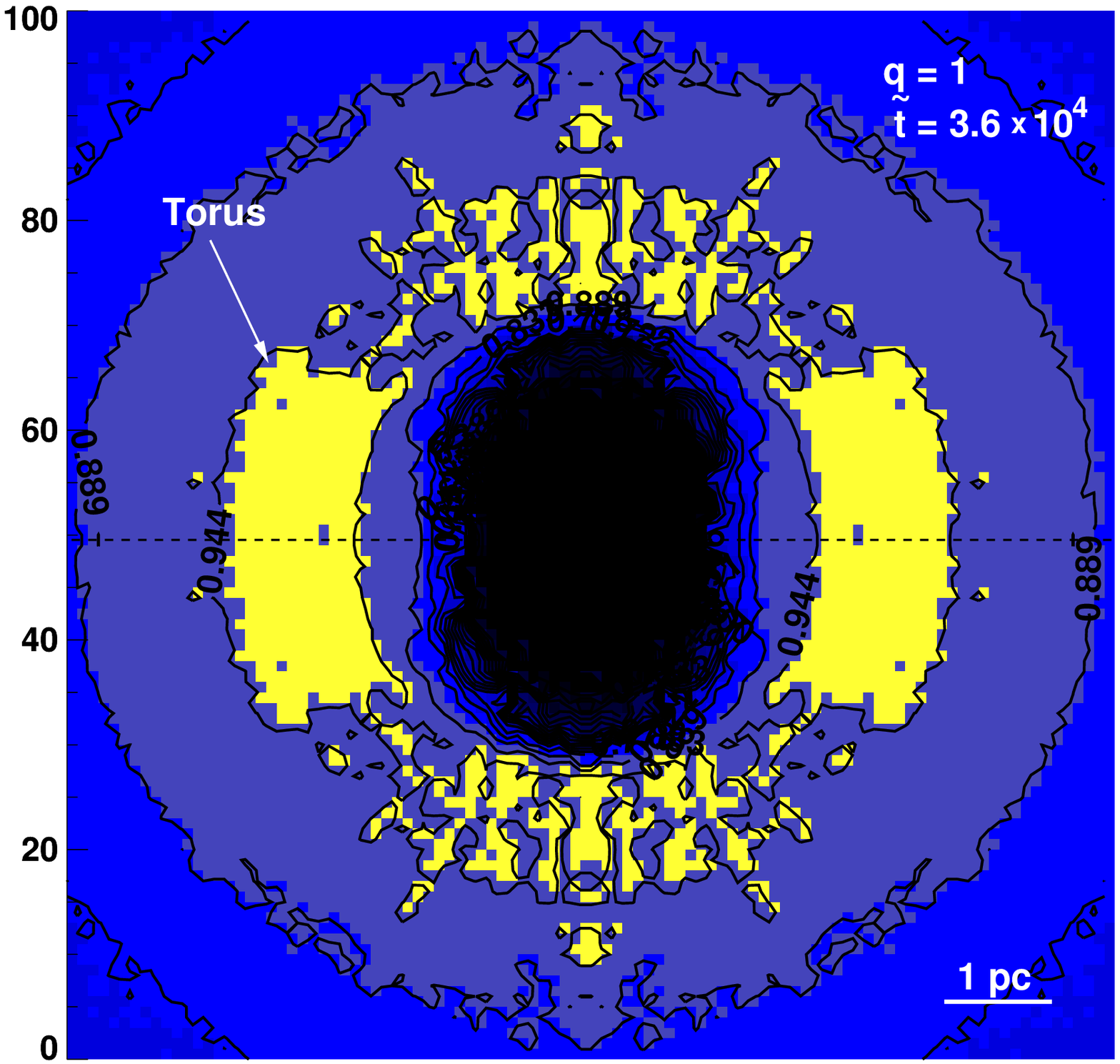}}%
\fi
\hspace{3mm}%
\ifpdf
  \resizebox{\picwidth}{!}{\includegraphics{f_gs_q1_t8_z_new.png}}%
\else
  \resizebox{\picwidth}{!}{\includegraphics{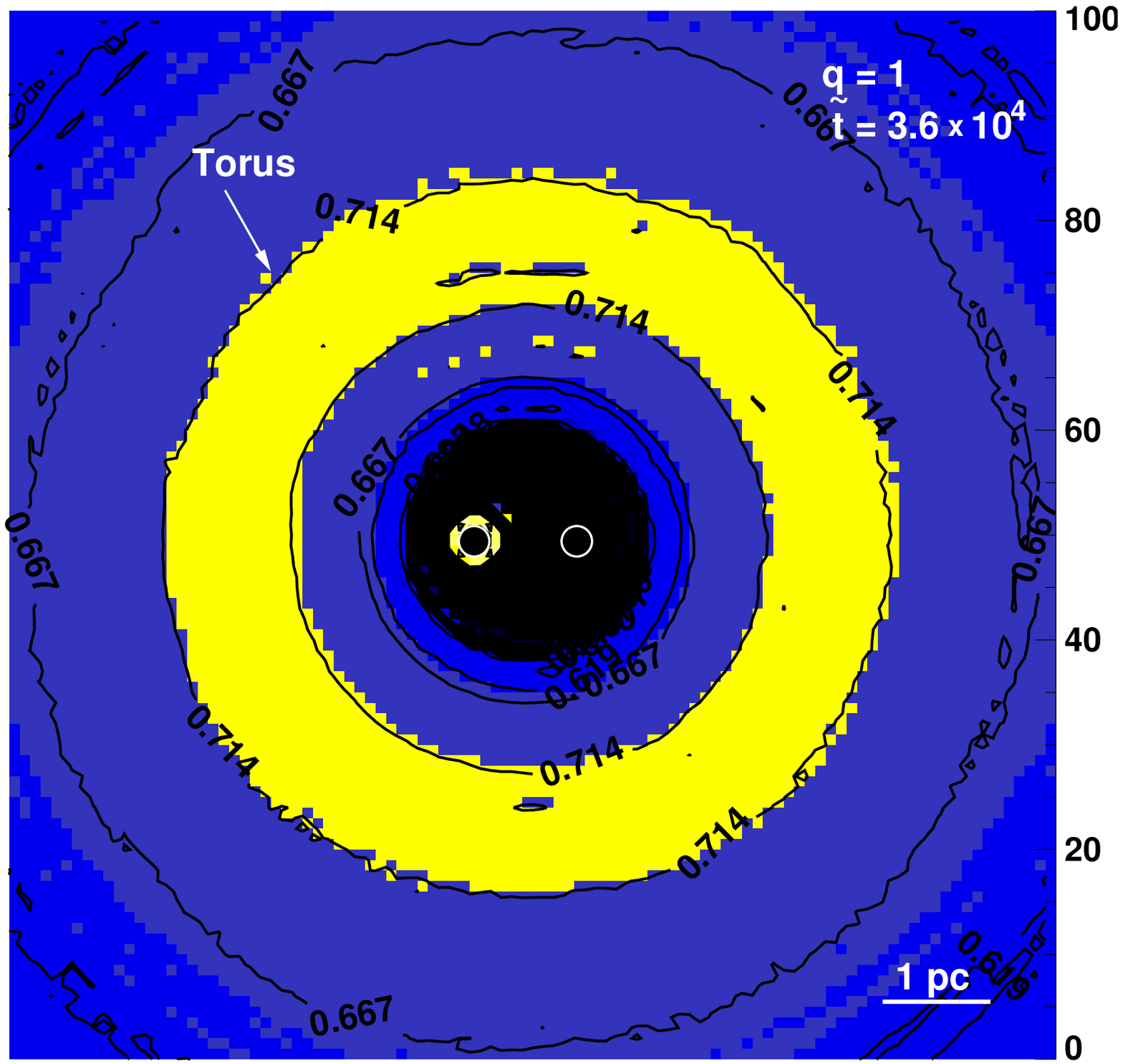}}%
\fi
\caption[]{
  Cuts through the stellar density in the comoving frame are shown
  with contours scaled logarithmically. The right panel displays the
  equatorial plane (BHs marked by black spots).  Perpendicular to it
  the $x=0$-plane is shown (left panel), with the $y$-axis drawn as
  dashed line, so that the BHs are in front and behind the
  paper-plane. The initial distribution is a gaussian and the
  mass-ratio is $1$. After initially stars close to the orbits in
  $0.5\ut{pc}$ distance from the center are ejected and the polar
  regions are depleted, finally a torus emerges in the equatorial
  plane at $r\sim 3\,{\rm pc}$. An expanded version of this diagram
  for the mass-ratio $10$ is shown in \p1.  }
\label{f_gs_cont_8}
\end{figure*}

In addition to these stars we will find in this region ($\tilde{r}\ge
50$) also those which happen to be close to their pericenters at the
very beginning of the simulation due to the choice of the initial
conditions.  As time proceeds they are moving on their orbits without
being noticeably affected by the secondary BH, whose perturbation to a
point-mass potential is negligible at large distances. Thus, keeping
their energy and angular momentum nearly conserved, these stars
diffuse into the outer regions ($\tilde{r}\ge 50$). The expansion of
the initial distribution is just a process of relaxation and would
have occurred also in a point-mass potential. Thus the region at
$\tilde{r}\ge 50$ is partly populated by stars which did not interact
with the binary.

Both these populations can be clearly distinguished in
Fig.~\ref{f_dx2_ldotr_q1}, where we plotted the angular momentum
versus the radius for each single star at the end of the simulation.
For $\tilde{r}\gtrsim 50$ one population is clustering around
$\tilde{L}\approx 2$ and the other having an upper limit at
$\tilde{L}\approx 10$, fading towards smaller angular momenta. Both
are extending to radii of about $10^4$, with the former population
being more pronounced at such big distances.

In a point-mass potential the energy reads in dimensionless units
\begin{equation}
\label{eq_energy}
\tilde{E}  = \frac{\tilde{v}_r^2}{2} +
\frac{\tilde{L}^2}{2\tilde{r}^2} - \frac{1}{\tilde{r}}\,,
\end{equation}
with the maximum energy of a bound star just below zero. To maximize
the angular momentum, all its kinetic energy is assumed to be stored
in tangential motion so that $\tilde{v}_r=0$.  Solving now for the
angular momentum gives
\begin{equation}
\tilde{L} < \sqrt{2 \tilde{r}}
\end{equation}
as the upper limit for bound stars. This power law with index
$-1/2$ is indicated in Fig.~\ref{f_dx2_ldotr_q1} by the dashed line.

However, the nature of the two populations at $\tilde{r}\gtrsim 50$
can be understood easily if we calculate the maximum possible angular
momentum for the stars as function of their initial pericenter. In
dimensionless form this is
\begin{equation}
\label{eq_rmin}
\tilde{r}_{-} = \frac{\tilde{L}^2}{1+\epsilon}\,.
\end{equation}
In order to obtain an upper limit for the angular momentum of the
diffused population, we assume that some stars at $\tilde{r}=50$ are
initially at their pericenter and move on highly eccentric orbits
($\epsilon\to 1$).  Solving then Eq. (\ref{eq_rmin}) for $\tilde{L}$
yields an upper limit of
\begin{equation}
\label{eq_llim}
\mli{\tilde{L}}{lim} = \sqrt{2\tilde{r}_{-}} = 10\,.
\end{equation}
If we use for the initial pericenter $\tilde{r}_{-} = 2$ instead of
$50$ we obtain $\mli{\tilde{L}}{lim} = 2$. This clearly shows that the
stars at $\tilde{r}\gtrsim 50$, whose angular momenta tend to $10$,
initially have been close to $\tilde{r}=50$ and thus diffused into
this region. On the other hand the stars with angular momenta
$\tilde{L}\approx 2$ had initial pericenters close to the radius of
the orbit of the secondary BH.  Here they strongly interacted with the
BBH before they were heated to distances up to $10^4$
(Fig.~\ref{f_dx2_ldotr_q1}). The heated stars in fact are very similar
to the ejected ones which have been discussed in detail in \p1.  Their
velocity distribution is also radially anisotropic
($\mli{\beta}{va}>0$, see Fig.~\ref{f_dx2_beta_q1}) and they are
moving on highly eccentric orbits. The heated stars just did not gain
sufficient energy as to escape from the binary.

\begin{figure}[t]
\ifpdf
  \resizebox{\picwidth}{!}{\includegraphics{f_dx2_vsq_q1.pdf}}%
\else
  \resizebox{\picwidth}{!}{\includegraphics{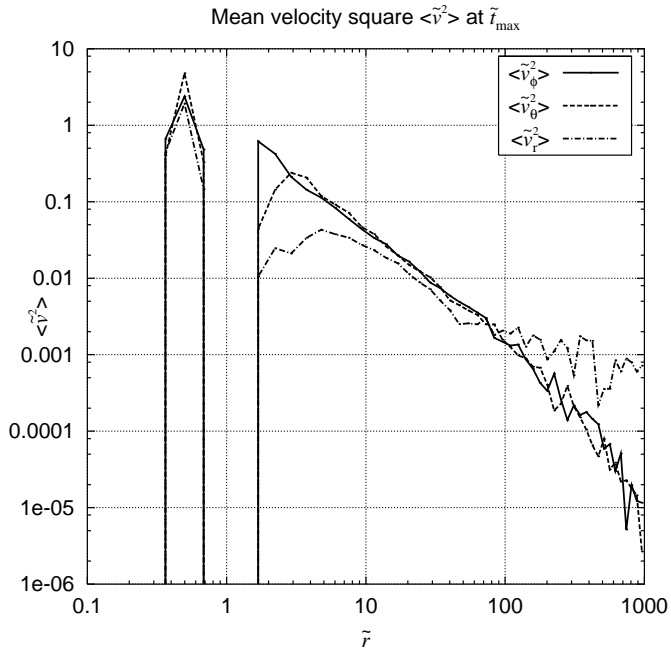}}%
\fi
\caption[]{
  The mean of the squares of the velocity components in spherical
  coordinates are displayed at $\mli{\tilde{t}}{max}$. In the range of
  the initial distribution ($\tilde{r}\le 50$) both tangential
  components exceed the radial one, while at larger distances it is
  just the opposite, in accordance with Fig.~\ref{f_dx2_beta_q1}. In
  the range from the inner edge of the torus at $\tilde{r}=1$ to the
  maximum of the number density at $\tilde{r}=3$ (\fgr{f_dx2_rho_q1}),
  the azimuthal component is larger than $\mean{\tilde{v}_\theta^2}$,
  in favour of the toroidal distribution of the stars. At larger
  distances these components become indistinguishable.}
\label{f_dx2_vsq_q1}
\end{figure}

This can be confirmed by computing the maximum distance a star can
travel, when it is moving at escape velocity, $\mli{\tilde{v}}{esc} =
\sqrt{2/\tilde{r}}$. If we integrate $\mli{\tilde{v}}{esc}$ with
respect to time from the beginning till the end of the simulation
($\mli{\tilde{t}}{max} = 5\times 10^5$) this distance is
\begin{equation}
\mli{\tilde{r}}{max} = \left(\frac{3}{\sqrt{2}}\, \tilde{t}_{\rm max}
  + \tilde{r}^{3/2}_{\rm i}\right)^{\frac{2}{3}} \approx 10^4\,,
\end{equation}
corresponding to the largest radii in Fig.~\ref{f_dx2_ldotr_q1}.  For
the initial radius $\tilde{r}_{\rm i}$ we have inserted values ranging
from $0$ to $50$, but since the first term in the brackets dominates
the expression it does not make a difference whether the stars started
in the center or from the edge of the initial distribution.

The bound stars in the intermediate range eventually assume a
torus-like structure as is shown in Fig.~\ref{f_gs_cont_8}, which is
taken from \p1. This plot displays cuts through the 3-dimensional
stellar density distribution in the comoving frame of binary after
about $10^{7}\,{\rm yr}$ when the BHs have merged.  The right panel
shows the equatorial plane ($z=0$) with the BHs marked by the black
points, $M_1$ to the right. The slice in the left panel is
perpendicular to the equatorial plane and contains the $x=0$-plane,
with the $y$-axis indicated by the dashed line. While the stars
initially have been distributed according to a gaussian, the central
parts have been scoured out and a torus in the equatorial plane of the
binary with a radius of about $3\,{\rm pc}$ is left. The basic
topology of the final density distribution can be understood in
physical terms, when we consider the torque exerted by the two black
holes on an orbiting star. Fig.~3 in \p1 shows of the normalized
torque the component $\dot{\tilde{L}}_{z,1}/\tilde{r}=\tilde{r}
\ddot{\phi}\sin^2(\theta)$ relative to the $z$-axis of the BBH as a
function of time for different angles $\theta$ between the rotation
axis of the BBH and the symmetry axis of the star's orbit. The bigger
the angle $\theta$ is (i.e. for orbits through the polar regions), the
more is the star's trajectory disturbed by the influence of the two
BHs. The cumulative effect of these large excursions in the polar cap
regions deplete the stellar population, leaving a torus behind.

The toroidal structure which the bound population assumes once the BHs
are merged is confirmed also by the kinematics of the stellar
distribution.  The tangential velocity components of a single star
moving on an orbit which is inclined by the angle $\theta$ to the
equatorial plane are
\begin{eqnarray*}
\tilde{v}_\phi & = & \mli{\tilde{v}}{t} \cos{\theta_v}\\
\tilde{v}_\theta & = & \mli{\tilde{v}}{t} \sin{\theta_v}\,,
\end{eqnarray*}
where $\theta_v$ is the angle enclosed by the binary's rotation axis
and the velocity-vector $\VEC{v}$ of the star. Thus the total
tangential velocity component is $\mli{\tilde{v}}{t} =
(\tilde{v}_\phi^2 + \tilde{v}_\theta^2)^{1/2}$. For stars constrained
to a torus-like volume, as shown in Fig.~\ref{f_gs_cont_8}, we expect
their azimuthal velocity component $\tilde{v}_\phi$ to exceed the
polar component $\tilde{v}_\theta$. This is verified in
Fig.~\ref{f_dx2_vsq_q1}, where we have plotted the mean square of each
velocity component versus the radius: At the inner edge of the torus
between 1 and $2\ut{pc}$ distance (Figs.~\ref{f_dx2_rho_q1} and
\ref{f_gs_cont_8}) the mean square of the azimuthal component is a
factor of about $10$ larger than $\mean{\tilde{v}_\theta^2}$. With
increasing radius both components approach each other and at about
$8\ut{pc}$ they are almost indisinguishable.

\section{Stellar winds comprising a patchy torus}
\label{sc_wind_cover}

Now that eventually the bound stars indeed constitute a torus-like
distribution which is geometrically thick, just as required by the
unification scheme, we may ask about the effects of this structure on
the radiation emitted from the center.  Of course the stars alone are
not able to obscure this radiation, since at a distance of about
$r=3\,{\rm pc}$ a number of order of at least $(r/R_\ast)^2 \approx
10^{12}$ stars would be required to achieve a covering factor of order
unity, using giant stars with a radius of $10^{11}\,{\rm m}$.  But if
there is a sufficiently large number of stars in the torus with dense
winds which absorb a significant fraction of the central radiation,
the complete torus might be opaque. To get an estimate of the optical
depth and density of the torus we have to examine its constituents,
the stars and their winds, more closely.

Solar type stars are not appropriate for the obsuration of the AGN,
since their winds are too weak. But there are various other stars
which can account for the obscuration and which we will call
`obscuring stars' (OS) in the following.  Using evolutionary tracks,
\citet{young77} showed that the total fraction on the giant branch in
a cluster ranges between 1\% and 2\% for an interval 1 to 5 billion
years after coeval star formation. In a first paper \citet{shull83}
assumed a dense stellar cluster to consist of two components, one
evolved component of red giants ($1\,M_\odot$, $100\,R_\odot$)
comprising 1\% of the stars, and a main-sequence component. In a later
paper \citep{voit88} the fraction of supergiants in a thermal cluster
was assumed to comprise only 0.01\%, giants ($1\,M_\odot$,
$10\,R_\odot$) 1\% and the rest of the cluster to be made up by
main-sequence stars. This would be only a small fraction with stars
having mass-loss rates of typically $10^{-5}\,M_\odot/{\rm yr}$.  On
the other hand the observed FWHM within the inner $160\,{\rm pc}$ of
both H$\alpha$ and [N II] and the equivalent width of the Ca II lines
are indicating the presence of an important population of red
supergiants in the nucleus of NGC 6951 \citep{perez00}.  The current
understanding of stellar winds is based on observations in the solar
neighbourhood and so far there are no results concerning the possible
effect of the intense radiation of the AGN on the wind structure. Thus
stars in AGN might be very different from stars in the outer galaxy
\citep{alexander_netzer94,alexander_netzer97}. There are theories that
the conditions in AGN are likely to enhance the mass-loss rate of
irradiated stars near the center
\citep{edwards80,shull83,voit88,tout89}. While this radiation might
not significantly increase the mass-loss rate \citep{voit88},
\citet{biermann89} and \citet{stecker91} suggested a significant
neutrino flux in AGN which may transform low-mass stars into red
giants \citep{macdonald91}.  Since we consider the the torus to be a
consequence of two merging galaxies with central massive black holes,
besides stars we also expect big amounts of gas to be brought to the
center, and therefore an enhanced formation rate of young stars with
high mass-loss rates.  This is supported by \citet{perez00}, who find
evidence that star formation occurs in bursts and continuously inwards
to the nucleus from a $5\,{\rm arcsec}$ radius in NGC 6951, and
\citet{oliva95}, who investigated red supergiants as starburst tracers
in galactic nuclei.  They find a large value of $L_{\rm H}/M$ in NGC
1068, confirming the presence of a decaying starburst in the central
regions ($R < 3''$).  We will condense the previous arguments into the
simple assumption, that the obscuring stars with strong winds amount
to a fraction of 1\% of the central cluster.

\begin{figure}
\ifpdf
  \resizebox{\picwidth}{!}{\includegraphics{tail_wind.pdf}}%
\else
  \resizebox{\picwidth}{!}{\includegraphics{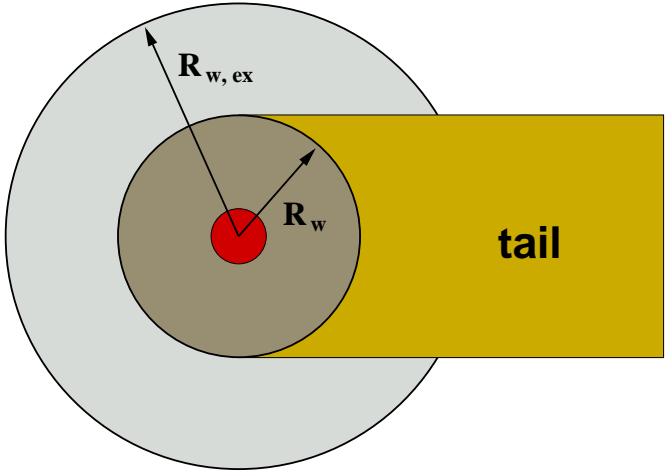}}%
\fi
\caption[]{
  This sketch illustrates how the radiation pressure of a strong
  source, located in a large distance $d$ ($d\gg \mli{R}{w,ex}$) left
  from the star, shapes the wind into a comet-like tail, pointing
  radially away from the source. The radius of the cross-section area
  of the wind is $\mli{R}{w}$, and $\mli{R}{w,ex}$ denotes the radius
  of a freely expanding wind.}
\label{f_wind_tail}
\end{figure}

The radiation pressure of a bright central source limits the
wind-radius in its extension and will shape it into an elongated tail,
pointing radially away from the central source. This is illustrated
for one star in \fgr{f_wind_tail}. The large light-grey shaded
circular area represents the freely expanding wind of the star in its
center.  The dark-grey area shows the smaller extension of the wind,
limited by a central source far left from the star, which shapes the
wind into a comet-like tail.

Such bending of the wind into a tail by radiation pressure can be
deduced from a crude and simple estmate. According to
\citet{mathis77}, Graphite with its evaporation temperature
$\mli{T}{evap}\sim 1500\ut{K}$ is the major contributor to the
extinction of the radiation outside the evaporation radius
$\mli{r}{evap}$. This corresponds to the distance from the center,
where the absorption of the central UV radiation by dust grains equals
the rate at which it is reradiated as thermal IR radiation, such that
the equilibrium temperature is the evaporation temperature. The
absorption efficiency of the grains in the IR is much smaller than in
the optical/UV, so that the reemission in the near IR (NIR) must be
optically thin. \citet{barvainis87} approximated the IR-absorption
efficiency by a powerlaw ($Q_\nu \propto \nu^\gamma$) with the index
in the range $1$ to $2$. This yields a spectrum with a considerably
narrower bump at $2\ut{\mu m}$ than is observed
\citep{sanders89b,haas00} and which corresponds to the optically thin
emission peak of the hottest grains at a temperature
$\mli{T}{gr}\approx 1500\ut{K}$. Hence, any model which seeks to
explain the NIR bump in terms of thermal dust emission requires a
range in the grain temperature. This is naturally included in our
model, since the heated dust will be in equilibrium at different
temperatures at different distances from the central radiation source.
For the grain temperature as function of the distance $r$ to the
center \citet{barvainis87} obtained
\begin{equation}
\label{eq_tevap}
\mli{T}{gr} = 1650\left(\frac{\mli{L}{uv, 46}}{\mli{r}{pc}^2}
\right)^{\frac{5}{28}} e^{-5\mli{\tau}{uv}/28} \,{\rm K}\,,
\end{equation}
with $\mli{\tau}{uv}$ as the optical depth of dust in the UV and
$\mli{L}{uv, 46}$ being the central UV luminosity in units of
$10^{46}\ut{erg/s}$. $\mli{r}{pc}$ is the distance normalized to
$1\ut{pc}$. Using $\mli{T}{evap}$ for the grain temperature and
solving this equation for $r$ gives the evaporation radius
\begin{equation}
\label{eq_revap}
\mli{r}{evap} = 4.05\,\mli{L}{uv, 46}^{1/2}
T_{{\rm evap},3}^{-14/5}\,{\rm pc}\,,
\end{equation}
where $T_{{\rm evap},3}$ denotes the evaporation temperature in units
of $1000\ut{K}$.

Assuming a star to be located inside $\mli{r}{evap}$ where the Thomson
cross-section $\mli{\sigma}{th}$ applies, its wind extension can be
calculated using the same Ansatz as for the Eddington limit, i.e.
equating the force of the central radiation acting on the wind
particles with the kinetic force of the wind,
\begin{equation}
\label{eq_windeq}
\frac{\mli{L}{uv} \mli{\sigma}{th}}{4\pi r^2 c} = \frac{\mli{m}{p}
\mli{v}{w}^2}{\mli{R}{w}}\,.
\end{equation}
With typical values of red giants for the wind velocity $\mli{v}{w}$
\citep{knapp85,winters94,hoefner96} this equation can be solved for
the extension of the wind,
\begin{equation}
\label{eq_est-wrad}
\mli{R}{w}\approx 1.2\times10^{-3} \left(\frac{r}{2\,{\rm pc}}\right)^2
\left(\frac{\mli{v}{w}}{10\,\frac{\rm km}{\rm s}}\right)^2
\left(\frac{\mli{L}{uv}}{10^{46}\,\frac{\rm erg}{\rm s}}\right)^{-1}\ut{pc}
\end{equation}

This is only a very rough estimate. In the following we are seeking
for a more robust value of the wind radius, which is based on
confirmed limits set by observations.  As discussed in Sect.
\ref{s_torus}, these tell us that the torus is optically thick and the
covering factor of its constituents, the stars, is about one.
Therefore the number density of obscuring stars, integrated along the
radial extension of the torus, yields a surface density of the number
of stars ($\mli{\Sigma}{os}$) which has to equal the inverse of the
cross-section area of the stellar winds, i.e.
\begin{equation}
\label{eq_windrad-rgdens}
\mli{\Sigma}{os} = \frac{1}{\pi \mli{R}{w}^2}\,.
\end{equation}
$\mli{R}{w}$ denotes the radius of the cross-section area of the
stellar wind perpendicular to its radial extension. To calculate the
surface density we employ the number density profile of the singular
isothermal sphere that we have already used in \p1 ($n\propto
r^{-2}$), and which has not been much altered by the merging-process
of the two BHs (\fgr{f_dx2_rho_q1}). With $n_0$ being the number
density in the distance $r_0=1\ut{pc}$ we can write
\begin{equation}
\label{eq_ndens}
n(r) = n_0 \left(\frac{r_0}{r}\right)^2\,.
\end{equation}
Integrating this density from the inner to the outer radius of the
torus ($\mli{r}{in}$ and $\mli{r}{out}$ respectively) gives the
surface density:
\begin{equation}
\label{eq_nsurfdens}
\mli{\Sigma}{os} = \int_{\mli{r}{in}}^{\mli{r}{out}} n(r)\d r =
\frac{n_0 r_0^2}{\mli{r}{in}}\left(1 -
  \frac{\mli{r}{in}}{\mli{r}{out}}\right)
\end{equation}

To determine the number density $n_0$, we have to calculate the number
of obscuring stars in the torus, which is the fraction $0.01$ of the
bound stars $\mli{N}{bn}$. Integrating the number density in
\eqn{eq_ndens} over the volume of the torus gives us the number of
obscuring stars $\mli{N}{os}$:
\begin{eqnarray}
\label{eq_Nos}
\mli{N}{os} & = & \int_0^{2\pi}\d\phi
\int_{\mli{\theta}{trs}}^{\pi-\mli{\theta}{trs}} \sin\theta\,\d\theta
\int_{\mli{r}{in}}^{\mli{r}{out}} n(r)r^2\d r \\
& = & 4\pi\cos\mli{\theta}{trs} n_0 r_0^2
(\mli{r}{out}-\mli{r}{in})\nonumber\,,
\end{eqnarray}
where $\mli{\theta}{trs}$ is the half-opening angle of the torus.
Solving this equation for $n_0$ and applying it to \eqn{eq_nsurfdens},
the wind radius is obtained with the help of \eqn{eq_windrad-rgdens},
\begin{equation}
\label{eq_wind-rad}
\mli{R}{w} =
\left(\frac{\mli{N}{os}}{4\cos\mli{\theta}{trs}\mli{r}{in}\mli{r}{out}}
\right)^{-\frac{1}{2}}\approx 1.8\times 10^{-3}\ut{pc}\,.
\end{equation}
For the parameters we assigned in the last step the values for the
case of a mass-ratio $q=1$: The inner radius is $\mli{r}{in} =
1\ut{pc}$ (see \p1 and Sect. \ref{s_torus}), and $5\ut{pc}$ is used
for the outer radius since the density drops quickly with increasing
distance.  The half-opening angle is $60^\circ$ (see
\fgr{f_gs_cont_8}) and $\mli{N}{os}=0.01\mli{N}{bn}$, with
$\mli{N}{bn}=3\times 10^8$ being a little more than the minimum of
stars required to allow the BHs to merge ($2.3\times 10^8$). The wind
radius computed in \eqn{eq_wind-rad} is in very good agreement with
our crude estimate in \eqn{eq_est-wrad} and thus confirms the
assumptions we made to obtain it.

Such a wind has to be optically thick ($\mli{N}{H}\simeq
10^{24}\ut{cm}^{-2}$) for all line of sights through its
cross-section.  We assume, that the density of such a tail, if
projected along its radial extension, yields a constant surface
density. Therefore the mass in the wind can be estimated to be of
order of
\begin{equation}
\label{eq_wind-mass}
\mli{M}{w} = \pi \mli{R}{w}^2 \mli{N}{H}\,\mli{m}{p}\approx 8.4\times
10^{-2}\,M_\odot\,.
\end{equation}

\begin{figure*}[t]
\hspace{1mm}
\ifpdf
  \resizebox{175mm}{!}{\includegraphics{f_star_wind.pdf}}%
\else
  \resizebox{175mm}{!}{\includegraphics{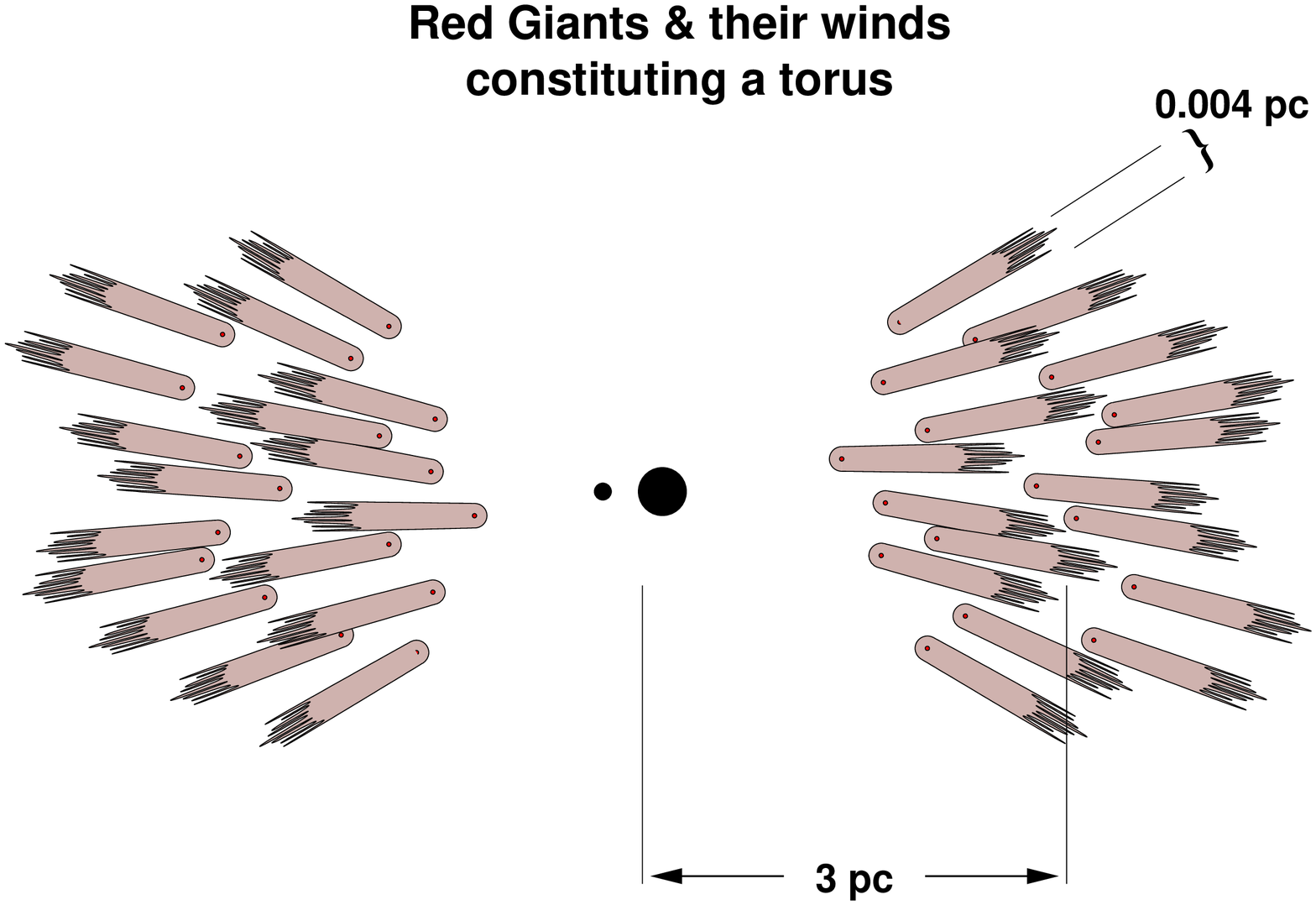}}%
\fi
\caption[]{
  This picture shows (not to scale) a sketch of the obscuring stars
  and their winds comprising a torus with the BBH in its center. In
  this cross-section through the torus, whose density peaks in about
  $3\,{\rm pc}$ distance, the obscuring stars are indicated by the
  small dots. They are surrounded by their winds, which are elongated
  by the central radiation into comet-like tails (shaded) with a
  lateral extent of about $0.004\,{\rm pc}$. The central radiation
  source with the BBH is marked by the two BH{}s in the center. An
  observer with the LOS close to the edge of the torus might see the
  center through a gap directly, while the view of an observer with
  his LOS close to the equatorial plane of the BH{}s is probably
  blocked by the winds of at least one obscuring star.  }
\label{f_star_wind}
\end{figure*}

The final picture of our model for the torus is illustrated (not to
scale) in \fgr{f_star_wind}. The sum of the winds of the obscuring
stars, shaped into elongated tails by the central radiation pressure,
form a patchy, optically and geometrically thick torus. For an
observer whose line of sight (LOS) is aligned with such a wind, the
nucleus is obscured and only radiation scattered into the LOS will be
detected. Since the number density of the stars decreases for smaller
inclination angles between the LOS and the symmetry axis of the torus,
the probability to observe the nucleus unobscured, or only partially
covered, through a gap in the torus, increases. Such gaps will be
continuously opend and closed due to the motion of the stars.
According to our prerequisites \eqn{eq_windrad-rgdens} is satisfied
and consequently the complete torus is optically thick on average.

With the number of obscuring stars and the mass contained in their
winds we can estimate the order of magnitute of the dust-mass which is
confined to the stellar winds outside the evaporation radius. The
relation between the dust and gas mass is given by
\begin{equation}
\mli{m}{d}\mli{n}{d} = 1.4 Z_{\rm d} \mli{m}{p}\mli{n}{H}\,,
\end{equation}
where $\mli{m}{d}$ and $\mli{m}{p}$ denote the mass of a dust grain
and the proton mass respectively. The number-densities are accordingly
given by $\mli{n}{d}$ and $\mli{n}{H}$ with $Z_{\rm d}$ as the
dust-to-gas mass-ratio. Integrating over the volume of the wind yields
$\mli{M}{d} = 1.4 Z_{\rm d} \mli{M}{gas}$. This is substituted in
$\mli{M}{w} = \mli{M}{d}+\mli{M}{gas}$ and solved for the dust mass:
\begin{equation}
\label{eq_winddust}
\mli{M}{d} = \frac{1.4 Z_{\rm d}}{1 + 1.4 Z_{\rm d}}\mli{M}{w} \approx
1.2\times 10^{-4}\,M_\odot\,. 
\end{equation}
In the last expression we used the wind mass of Eq.
(\ref{eq_wind-mass}) and for the dust-to-gas mass-ratio $Z_{\rm
  d}\approx 10^{-3}$ \citep{hoefner96}.  Multiplying this dust mass
with the number of obscuring stars needed to maintain a covering
factor of 1 ($\mli{N}{os} = 3\times 10^6$) gives the total dust mass
contained in the wind of the obscuring stars,
\begin{equation}
\label{eq_totaldust}
\mli{M}{d, total} = \mli{M}{d} \mli{N}{os} = 350\, M_\odot.
\end{equation}
This is about a factor of $10$ more than the minimum values given by
\citet{sanders89b}, which are required to fit the spectra.  Within
spheres of the radii $1\,{\rm pc}$ and $10\,{\rm pc}$ they obtain for
the minumum of dust amounts $0.2\,M_\odot$ and $20\,M_\odot$
respectively, using for the grains the parameters given in
\citet{mathis77} and \citet{biermann80}. If all grains within the
given radius are larger than $\sim 1\ut{\mu m}$, or are not directly
exposed to UV radiation, the required dust mass is $\sim 10$ times
higher, in good agreement with \eqn{eq_totaldust}. Such shielding of
the outer regions of the elongated winds by the inner parts is a
natural consequence of our model and diminishes the UV irradiation of
the outer parts of the wind. Also, due to a covering factor of about 1
it is likely that the wind of a star is covered by that of another one
at smaller distances to the central source.

The picture which emerges from our simulations and the results is the
following: If a BBH is surrounded by a stellar population of
comparable mass, the stars which are ejected in violent interactions
with the BH{}s are able to carry away enough angular momentum of the
binary so that it hardens till finally gravitational radiation
dominates the shrinking process of the BH{}s (\p1). These eventually
coalesce after about $10^7\,{\rm yr}$, having spent most of the time
in the range when the ejection of stars dominates the hardening.  On
the other hand for the same number of initial stars there are enough
left which are bound to the binary and form a torus-like distribution
which peaks in about $3\,{\rm pc}$ distance.  In this section we could
show that the amount of obscuring stars in the remaining cluster is
sufficiently large so that their winds achieve a total covering factor
of about 1. These winds are are optically thick. Thus the patchy torus
as a total is geometrically as well as optically thick so that our
model can indeed account for a dusty torus in the center of AGN, as is
demanded by the unification model.

\subsection{Relativistic boosting}
\label{sc_boosting}

Within the solid angle of a relativistc jet, the radiation is boosted
and so the flux density of the approaching radiation, seen under an
angle $\theta$ to the velocity of the moving source, scales as
\citep{rybicki79,longair81}
\begin{equation*}
S = D^{3-\alpha} S',
\end{equation*}
with $\alpha$ being the spectral index and $D$ the relativistic
Doppler factor
\begin{equation*}
D = \frac{1}{\gamma (1 - \beta \cos \theta)}\,.
\end{equation*}
The primed quantities refer to the comoving frame of the source. Thus,
for an index $\alpha \sim -1$, the observed flux density depends on
the Doppler factor to the power of four. Half of the radiation is
emitted in a cone of half-opening angle $\theta \simeq 1/\gamma$, if
$\gamma \gg 1$, and thus is strongly beamed. In this limit ($\gamma\gg
1$) we have $D\approx\gamma$ and the luminosity in the cone is a
strongly increasing function of $\gamma$. If we assume the luminosity
$L$ to scale in the same way as $S$ with $\gamma$, it increases very
strongly in the beaming cone (i.e. by a factor of $10^4$ for $\gamma =
10$ and $\alpha = -1$). Consequently, according to \eqn{eq_est-wrad},
the extension of the winds of the obscuring stars, exposed to the
strong beamed radiation, is negligible and the covering factor in this
region tends to zero.  This means that within the beaming cone, and if
the jet is precessing, within the cone of precession, the emission
from the jet frees the polar cap regions from obscuring clouds, and
therefore supports the toroidal structure of the central absorber.

\subsection{Lifetime of the torus}
\label{sc_lifetime}

After the merger of the BBH is completed, the toroidal structure of
the stellar distribution will not collapse and is stable (\p1).
Therefore the torus will decay on timescales of the lifetime of its
constituents, the obscuring stars (i.e. red super-giants and bloated
super-giant stars) or has to be replenished with stars from outside.
Because after the merger no torque acts anymore on the stars, the
stars fed to the torus from larger distances would have to be accreted
in an axi-symmetric way to maintain the torus-like shape of the
circum-nuclear stellar distribution. The jet-outflow probably helps to
keep the polar-cap regions free from stars. But probably the rate of
stars accreted to the torus will decrease with time, as the outer
regions, stirred up by the merger, will relax with time.

Hence, unless the required number of obscuring stars is not maintained
for a longer time by the evolution of the other stars in the cluster,
the life-time of the torus will be of comparable order as the time
needed by the BHs to coalesce from the distance where a torus emerged
due to their torque acting on the surrounding stars. Once the binary
has become hard ($a \sim 1\ut{pc}$), the BHs merge on scales of about
$10^7\,{\rm yr}$ (\p1).

In the next sections we present implications and predictions of the
proposed model.

\begin{figure*}[t]
\hspace{2mm}
\ifpdf
  \resizebox{178mm}{!}{\includegraphics{spin_paper2.pdf}}%
\else
  \resizebox{178mm}{!}{\includegraphics{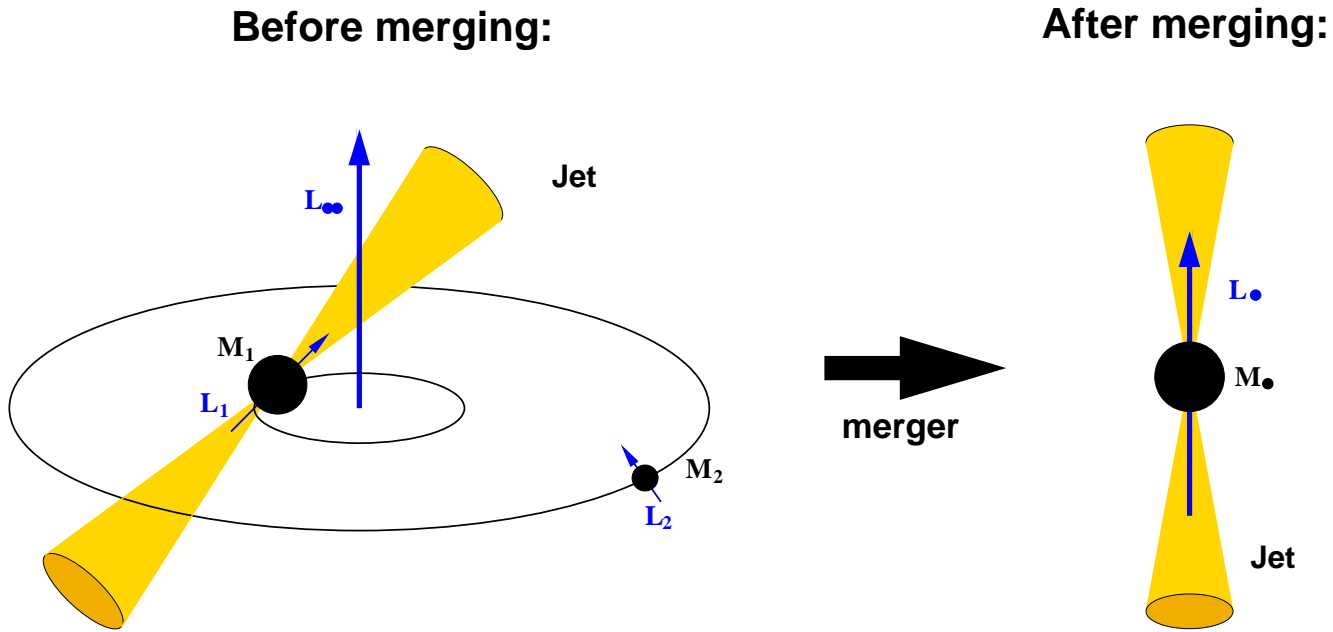}}%
\fi
\caption[]{
  This figure illustrates the change of the direction of the spin of
  the BH, induced by the merger of 2 massive BHs, and consequently the
  change of the direction of the jet. The left panel shows the
  situation before the merger, when the jet is aligned with the
  individual spin of the primary black hole of the binary system. The
  orbital angular momentum $L_{\BHBidx}$ of the BBH and the spins of
  both BHs ($L_1$ and $L_2$) are randomly orientated to each other,
  since no direction of the plane of the merging galaxies is preferred
  relative to the spin of both the BHs in their centers. After the BHs
  coalesced (right panel), the different spins combine to that of the
  merged BH with a new direction ($L_{\bullet}$). The jet emanating
  from the merged BH will be aligned with the new spin, which will be
  dominated by $L_{\BHBidx}$, and therefore has to jump form its
  orientation before the merger ($L_1$) into this new direction.  }
\label{f_spin_sketch}
\end{figure*}

\section{Jet-flip due to spin-flip of the primary BH}
\label{sc_jet-flip}

As we have already mentioned in \p1, we expect the merger of two
galaxies of comparable size to result in a galaxy of early type, i.e.
of elliptical shape (e.g. \citet{toomre77}, \citet{biermann79},
\citet{farouki82}). The spiral structure the parent galaxies might
have had will not survive such an event. On the other hand, radio-loud
AGN so far have not been found in spiral host-galaxies, what suggests
radio-loud galaxies to be the product of one or more major mergers.
Since observations show that spectra from AGN in the range from IR to
X-rays, dominated by emission from an accretion disk, are basically
the same \citep{sanders89b}, \citet{wilson95} argue that the
distinction between radio-loud and radio-quiet AGN is due to the
different spins of the central supermassive BHs. They propose
radio-loud objects to harbour a fast spinning super-massive BH as
result of a recent major merger, whith the assumption of non-rotating
BHs in the parent galaxies. The time scales we obtained for the
merging of the BHs are considerably smaller than a Hubble-time, in
agreement with such a model.

But what happens, if one of the progenitor galaxies has a fast
spinning BH in its center with a powerful radio-jet aligned with it's
spin? The situation before such BHs merge is depicted in the left
panel of Fig.~\ref{f_spin_sketch}. The spins of both BHs and the
orbital angular momentum of the BBH are randomly oriented to each
other, because the merger of both galaxies does not proceed in a
favored plane relative to that of the galaxies. While the secondary
BHs spin is assumed to be negligible, we adopt for the primary BH a
spin that is sufficiently large to power a strong radio-jet according
to the spin-paradigm ($L_1 \gg L_2$). Its upper limit is given by $M_1
c R_{\rm g} = G M_1^2 /c$, with $R_{\rm g}$ being the gravitational
radius.  For the orbital angular momentum we have
$L_{\BHBidx}=M_1\sqrt{GM_1 a}/\sqrt{q (1+q)}$, where $q=M_1/M_2 \ge 1$
is the mass-ratio of both BHs and $a$ is the semi-major axis.  Thus
the ratio of these two angular momenta is
\begin{equation}
\label{eq_spinratio}
\frac{L_{\BHBidx}}{L_1} = \frac{1}{\sqrt{q
    (1+q)}}\sqrt{\frac{a}{R_{\rm g}}}\,.
\end{equation}
It is a very interesting question now, what actually dominates the
final spin of the merged black hole, $L_\bullet$. When the distance of
both BHs has shrunk to about the last stable orbit ($R_{\rm ls} = 6
R_{\rm g}$ in case of zero spin, and in case of a Kerr BH it depends
on the inclination, see \citet{chirvasa02}) due to the emission of
gravitational radiation, general relativistic effects become
important. At this distance the orbital angular momentum dominates
over the maximum possible spin of $M_1$, up to a mass-ratio of $q=2$.
This is actually a conservative limit since we neglected the last
stable orbit around $M_2$. Hence the spin of the merged BH might be
dominated by $L_{\BHBidx}$ rather than $L_1$, with the merged black
hole $M_\bullet$ spinning close to its maximum value $L_\bullet = G
M_\bullet^2 /c$. This situation is shown in the right panel of
Fig.~\ref{f_spin_sketch}. A new powerful radio-jet is emanating from
the center, aligned with the BH's spin which is pointing in the
direction of $L_{\BHBidx}$, and therefore parallel to the
symmetry-axis of the torus. This means that the jet flips into a new
direction. The angle enclosed by the old and new jet-axis corresponds
to that between $L_1$ and $L_{\BHBidx}$ and hence can assume any
value. After the merger the old jet is not fed anymore and its lobes
will slowly fade away, while the new jet is digging its way through
the ambient medium and the polar caps of the torus.

This might be the explanation for some of the most peculiar objects
seen on the sky, the X-shaped radio-galaxies. They show a secondary,
presently non-active pair of lobes, which is larger than and sometimes
almost perpendicular to the presently active pair
\citep{parma85,rottmann98,rottmann01}, which we interpret in this
picture as the new jet. This scenario is also in agreement with the
involved time scales of some $10^7\,{\rm yr}$ for the merger and $\sim
6\times10^7\,{\rm yr}$ in B2 0828+32 for the spectral aging of the
secondary lobes (\citep{rottmann01}, priv. comm. H. Rottmann).

Such a major merger will also increase the accretion-rate onto the BH
which might be in favour of the arguments of \citet{meier01}, who
suggests an association of jet production with geometrically thick
accretion flows and BH rotation.

For larger mass-ratios ($q > 2$) the possible influence of
$L_{\BHBidx}$ on the final BH spin is decreasing, and therefore also
the bending of the jet into a new direction will have a smaller
amplitude. Thus we do not expect a jump of the jet in minor mergers
where $M_2 \ll M_1$ and therefore $L_{\BHBidx} \ll L_1$ (see
Eq.~(\ref{eq_spinratio})) and ascribe the phenomenon of X-shaped
radio-galaxies to recent major mergers, where one of the progenitors
already had a powerful radio-jet. If radio strength is correlated with
the spin of the BH, then the observations imply that a BH merger of
spinning black holes leaves the ratio of spin to mass $J/M$ large.

Before the two BHs finally coalesce, the surrounding patchy torus will
have emerged on time scales of $10^7\,{\rm yr}$ (\p1) in the plane of
the merger, with its symmetry axis pointing along the binary's angular
momentum. As long as the semimajor axis of the BBH is much larger than
the last stable orbit of $M_1$, the old jet of the primary AGN is
unaffected by $M_2$ and still fed, if the feeding mechanism from the
accretion disk is not interrupted.  Because the orientation of this
jet is not correlated with that of the torus, it might flow into the
solid angle covered by the torus, where it will interact with stellar
winds and ISM in between. After the BHs coalesced the young
past-merger jet is digging its way through the polar cap region of the
torus, flowing along its symmetry axis. So both jets, the old and new,
are interacting on sub-${\rm kpc}$ scales with the ambient clumpy
medium producing strong radio emission. This might be the cause of the
so called Compact Symmetric Objects (CSO), which is a class of
powerful radio sources consisting of high unbeamed luminosity radio
emission regions separated by less than $1\,{\rm kpc}$ and situated
symmetrically about the center of activity
\citep{phillips82,wilkinson94}. It is thought that the high-brightness
regions are due to hot spots and mini-lobes, which are created by the
termination of jets streaming out into opposite directions from the
center, see \citet{owsianik98b} and references therein. Instantaneous
speed variations of some components in CSO 0710+439 are most simply
explained by interactions with a dense cloud medium, while other
components are rapidly advancing through an intercloud medium.  The
expansion velocities are all of order $0.2\, h^{-1}c$ and therefore
the age of these sources is estimated to be a few thousand years,
showing that they are young rapidly growing sources
\citep{owsianik98a,owsianik99}. It is concluded that CSO{}s are
probably young extragalactic radio-sources which will evolve via
Medium-size Symmetric Objects into Large-size Symmetric Objects such
as lower luminosity FR~II double radio-sources
\citep{fanti95,readhead96,owsianik98b}.

If the X-shaped radio galaxies are correctly explained with our
proposed picture, the resulting gravitational wave pattern will
strongly depend on the orientation of the three relevant spins
\citep{chirvasa02} and therefore on the re-orientation of the primary
black hole. It experiences a spin-flip in this picture.

\section{A patchy torus -- explanation for BAL~QSOs and transition
from Type 1 to 2}
\label{sc_bal-trans}

The broad absorption line (BAL) quasars comprise about $10\%$ of the
optically selected quasars (i.e. \citet{antonucci01} or
\citet{schmidt99}), but since they are hard to find at optical
wavelengths, their real fraction is thought to be in the range $20\% -
30\%$. Moreover the BAL covering factor is thought to be much less
than unity and consequently there must be many objects, if not all,
being intrinsically the same as BAL quasars, but not classified as
such if seen from other directions. The region which is responsible
for the formation of the BAL partially absorbs the broad emission
lines, and therefore it must be outside the BLR. Also the polarization
in the broad emission lines is observed to be lower than the
polarization in the absorption troughs of the BALs.

Inside the torus, with its inner edge close to the evaporation radius
of dust ($\sim 1\,{\rm pc}$) the heated matter will be accelerated and
streams away from the nucleus through the gap between jet and torus
either in form of clouds or as a wind. Here, above the surface where
the density of the torus gradually decreases, the clouds are exposed
to more intense radiation.  These arguments fit well into the
unification model, where the BAL~QSOs are quasars seen at intermediate
inclination angles, with the line of sight grazing the surface of the
obscuring torus \citep{weymann91,voit93} and thus is supported by our
model.

With the assumption that the fraction of BAL~QSOs of $\chi=0.2$ to
$0.3$ represents the covering factor $C_{\rm BAL}$ in the same range
\citep{weymann97}, we can estimate the solid angle they comprise. We
just have to solve the equation of the covering factors, $\chi =
\mli{C}{BAL}/\mli{C}{QSO}$, for the polar angle $\mli{\theta}{BAL}$,
using the relations $\mli{C}{QSO} = 1 - \cos \mli{\theta}{trs}$ and
$\mli{C}{BAL} = \cos \mli{\theta}{BAL} - \cos \mli{\theta}{trs}$. In
this notation the QSO exhibits BAL features if seen under inclination
angles in the range $\mli{\theta}{BAL} \lesssim \mli{\theta}{incl}
\lesssim \mli{\theta}{trs}$. Assuming the half-opening angle of the
torus to be $\mli{\theta}{trs} = 45^{\circ}$ and using the above
fractions of BAL~QSOs $\chi =0.2$ and $0.3$ we obtain
$\mli{\theta}{BAL} = 40^{\circ}$ and $37^{\circ}$ respectively.  Hence
for an intermediate inclination angle in the range of about
\[
38^\circ \lesssim \mli{\theta}{incl} \lesssim 45^\circ
\]
the observed quasar will appear as a BAL~QSO, in agreement with
\citet{schmidt99} who find BAL~QSOs to be typically 2.4 times more
polarized in the optical than QSOs.

Among the BAL~QSOs the subclass of low-ionization absorbers shows
evidence to be stronger polarized and to be reddened by dust
\citep{sprayberry92,egami96}.  \citet{green01} assumed that the same
intrinsic power-law with index $\Gamma\approx 1.8$, which is
consistent with the mean slope of radio quiet QSOs and has been
derived for high-ionization BAL~QSOs, applies to all BAL~QSOs and is
partially covered. This assumption has been confirmed in some recent
publications \citep{gallagher01a,green01,gallagher01b}.  For
low-ionization BAL~QSOs, which are weak in X-rays, \citet{green01}
infer that they are enshrouded by an additional intrinsic column
density of nearly $10^{23-24}\,{\rm cm}^{-2}$. These low-ionization
BAL~QSOs have been proposed by \citet{brotherton97} as the most
edge-on QSOs.

For larger inclinations than the intermediate range the line of sight
will intersect with more absorbing material since it approaches the
surface of the patchy torus, which does not have a sharp defined edge.
Consequently less of the BLR is seen directly and the fraction of
scattered light as well as the polarization is increasing. With
increasing inclination the view to the nucleus is increasingly blocked
by the torus and the quasar will appear more reddened (IRAS QSO) till
finally, if seen edge on, it will appear as a QSO of Type 2 (ULIRG or
FR~II radio galaxy). On the other hand, objects seen at lower
inclinations don't suffer absorption of the central radiation by the
torus and are classified as quasars.

In the special case of the quasar FIRST J101614.3+520916
\citet{gregg00} find BAL features in the optical and radio loud
emission from lobes of classic FR~II type at the same time.  According
to the authors this quasar has some properties in common with
low-ionization BAL~QSOs, as for example higher reddening. They deduce
an inclination of the jet-axis to the LOS of more than $40^{\circ}$
and conclude that this quasar is in contradiction with the orientation
model as explanation for BAL QSOs. \citet{gregg00} suggest that this
quasar rather happens to be seen in a rare and short-lived state,
showing both, BAL features and developed radio lobes at the same time,
and postulation that it is a rejuvenated quasar, possibly through
merger. Contrary to this conclusion we claim that these observations
actually support the orientation model. According to the above
numbers, BAL~QSOs are seen at intermediate inclination angles in the
range between $\sim 38^{\circ}$ and $\sim 45^{\circ}$, in very good
agreement with the inclination obtained for this quasar. Thus the LOS
to the core of J101614.3+520916 is grazing the surface of the patchy
torus, whose constituting absorbers, the stellar winds, are increasing
in number density towards the equatorial plane.  At higher
inclinations the torus would block the free view to the center
completely and no BALs could be detected, so that this object would
appear as a typical FR~II quasar. This is also in agreement with
BAL~QSOs, that show large radio lobes, being so rare.  We do not
exclude evolutionary effects, but as shown above, we also expect a
merger to seriously affect the jet, what should be detectable. And the
question to be answered for the evolutionary/rejuvenated scenario is:
What is the explanation then for the difference between the
radio-quiet BAL and non-BAL quasars?

Due to the patchy nature of the torus, we propose partial covering
absorption plus scattering in order to fit the spectra of BAL~QSOs.
Another important consequence of the patchy torus is, when the
absorbing clouds are moving through the line of sight, the column
density changes and variations in the spectrum are expected. This is
then a natural explanation for the variability of the BAL quasar
PG~2112+059 between the \emph{ROSAT} and \emph{ASCA} observations on
scales of $6\,{\rm yr}$, being due to changes in the absorber, either
the ionization parameter, or the column density. We investigate such
variabilities in more detail in the following.


One observable consequence of an obscuring patchy torus, as developed
in this paper, is a possible variation of the absorption properties,
i.e. the column density.  A covering factor of the torus of about
unity means that not necessarily all lines of sight through the torus
to the central radiation source are covered and that due to the
motions of the stars in the torus under the influence of the central
BHs such gaps will form and will be closed continuously.

When an obscuring cloud or stellar wind of the torus is moving into
(out of) the line of sight of the observer to the radiation source,
the column density will drastically increase (decrease) and the source
will gradually appear much weaker (brighter) in luminosity. For our
chosen AGN with $M_1 = 10^8 \,M_\odot$ we obtained the circumnuclear
torus to be located in a distance of a few parsecs to the center. The
radius of the wind of the obscuring stars we computed to be of order
of $0.002\,{\rm pc}$. For a star moving on Keplerian orbits in the
potential of such a BH, the velocity is $v = \sqrt{G M_1 /r}$.
Therefore the time required for the star's wind to move along its
diameter from one edge to the other through the line of sight is
\begin{equation}
\label{eq_tobscure}
t_{\rm var} = 2\frac{R_{\rm w}}{v} \approx 10
\;\frac{R_{\rm w}}{0.002\,{\rm pc}} \left(\frac{r}{{\rm
     3 pc}}\right)^{\frac{1}{2}} \left(\frac{M_1}{10^8
    \,M_\odot}\right)^{-\frac{1}{2}} \,{\rm yr}\,.
\end{equation}
Consequently variations of the column density and the luminosity are
expected to happen on scales of a decade. Since the density of stars
is enhanced towards the equatorial plane of the torus it is likely
that there is more than one star in the line of sight and thus a
transition from optically thick to thin or vice versa will be a rare
event in an edge-on seen torus. But if the LOS grazes the surface the
torus the chance to observe such a transition will be higher. And
indeed there are a couple of observations which fit well to this
interpretation. Before comparing the above computed time scale for a
transformation from Type~1 $\to$ 2 with observations, we want to scale
it just in a simple way with the mass of the central BH.

Assuming the luminosity to be proportional to the eddington, $L_{\rm
  edd} = 4\pi G c M_1 m_{\rm p} /\sigma_{\rm Th}$, we have $L\propto
M_1$. Thus with Eq.~(\ref{eq_est-wrad}) the wind radius scales as
\[
R_{\rm w}\propto \frac{r^2}{L}\propto \frac{r^2}{M_1}\,.
\]
Together with $v_{\rm kep}\propto \sqrt{M_1/r}$ the dependency of the
time on mass and radius is:
\[
t_{\rm var}\propto \frac{r^{5/2}}{M_1^{3/2}}\,.
\]
There are now at least two possibilities for the choice of the inner
radius of the torus:

(1) The evaporation radius, for which we have $r_{\rm evap}\propto
L^{1/2}\propto M_1^{1/2}$ (see Eq.~(\ref{eq_revap})).  Thus we obtain
the dependency
\begin{equation}
t_{\rm var}\propto M_1^{-1/4}\,,
\end{equation}
what means that the time required for a cloud passing through the line
of sight is decreasing with increasing mass of the central BH.

(2) Another choice for the radius of the torus is the distance of the
BHs, when the binary is hard, since the simulation showed that this
coincides to within a factor of a few with the radius of the torus
(\p1).  According to \citet{milos01} the binary becomes hard at a
semi-major axis $a_h = G M_{\BHBidx}/8\sigma^2$. Using the relation
\[
M_\bullet = 1.3\times 10^8\,M_\odot \left(\frac{\sigma}{200\,{\rm
km/s}}\right)^\alpha\,,
\]
with $\alpha = 4.72 (\pm 0.36)$ (found by Meritt \& Ferrarese
\citeyear{merritt01a,merritt01b}) we get
\begin{equation}
t_{\rm var}\propto M_1^{(\alpha -5)/\alpha} \approx M_1^{-8/135}\,.
\end{equation}
Again there is a negative exponent, albeit small, so that $t$ is
decreasing with increasing $M_1$, as for $r=r_{\rm evap}$.

Since the exponents in the time-mass relation are much smaller than
$1$ for both choices of the inner radius of the torus, the
transition-time does not depend strongly on the mass of the central
BH, and is almost constant on orders of $10\ut{yr}$. For a patchy
absorber it is possible that more than one cloud or stellar wind
happen to block the line of sight, and depending on their individual
velocities and directions in which they move, it is possible that the
decrease and increase of the central flux proceeds on different time
scales.

In the following we will give some examples of sources which are in
agreement with our patchy torus model within the unification scheme
and show the expected variations in column density on scales of a
decade.

\emph{NGC~7582:} Compared to the \emph{EXOSAT} observations in 1983
\citet{warwick93} detect $4\,{\rm yr}$ later with \emph{Ginga} a
significant increase of the column density to $\sim 4.6\times
10^{23}\,{\rm cm}^{-2}$ by a factor of about $3$. While the spectral
index between the two \emph{ASCA} observations from November 1994 and
1996 does not show a significant variation the column density seems to
have increased by $\sim 44\%$ \citep{xue98}. In the hard and soft
X-ray band of the \emph{BeppoSAX} observation in November 1998
\citet{turner00} find the nucleus clearly to have brightened since the
1994 \emph{ASCA} observation. While this increase seems to be
consistent with the appearance of holes in the full screen ($1.4\times
10^{23}\,{\rm cm}^{-2}$) the authors also find evidence for larger
absorption at the \emph{BeppoSAX} epoch with a ``thick absorber''
covering 60\% of the nucleus, corresponding to a half-opening angle of
a torus of $\sim 53^\circ$, which has a column density of $1.6\times
10^{24}\,{\rm cm}^{-2}$. Both these values are in good agreement with
the column density we computed for the torus and that what is obtained
if the matter would be smoothly distributed in the volume of the torus
(see \eqn{eq_nhsmooth} in Sect. \ref{sc_massratio}).  The total
timescale of about $15\,{\rm yr}$ for major variations in the column
density is in very good agreement with our predictions.

\emph{NGC~2992:} This source has been monitored over a time of
$20\,{\rm yr}$ with different telescopes (\emph{HEAO}
\citep{mushotzky82,singh85}, \emph{Einstein}
\citep{halpern82,turner91} \emph{EXOSAT}, \citep{turner89},
\emph{Ginga} \citep{nandra94}, \emph{ASCA} \citep{weaver96} and
\emph{BeppoSAX} \citep{gilli00}).  During $16\,{\rm yr}$ the
$2-10\,{\rm keV}$ flux declines steadily by a factor of $20$ and is
$4\,{\rm yr}$ later back to its initial value measured in $1978$. Over
this time range the spectrum retains a constant shape, which is only
different in the low state observed by \emph{ASCA} in $1994$, when
there are also no short-time variabilities are seen. This is actually
what is expected, since according to our idea in this stage the direct
view to the center is blocked by a cloud.

\emph{NGC~4051:} While the flux in the range $2-10\,{\rm keV}$ of this
Seyfert~1 galaxy has been almost constant for more than $10\,{\rm yr}$
till the \emph{RXTE} observations by \citet{uttley98} it is measured
to be 20 times fainter $1.5\,{\rm yr}$ later with \emph{BeppoSAX} in
1998 by \citet{guainazzi98}. The time scale of $\sim 1.5\ut{yr}$ for
the decrease of the luminosity by a factor of $20$ might be too short
to be accounted for by a cloud of the patchy torus moving into the
line of sight.

\emph{NGC~1365:} Between the \emph{ASCA} observations and the
\emph{BeppoSAX} in August 1997 three years later the flux in the range
$2-10\,{\rm keV}$ has increased by a factor of $6$ \citep{risaliti00}.

\emph{NGC~3227:} In their comparision of the data from \emph{ASCA}
observations in 1993 and 1995 and \emph{ROSAT} observations in 1993
\citet{george98} show that the column density of the ionized absorber
increased by about an order of magnitude, i.e. from $N_{\rm H}\sim
3\times 10^{21}\,{\rm cm}^{-2}$ in 1993 to $N_{\rm H}\sim 3\times
10^{22}\,{\rm cm}^{-2}$ in 1995. They think this to be most naturally
explained by a cloud of material moving into the cylinder of sight.

\section{Influence and implications of the binary's mass-ratio}
\label{sc_massratio}

Our simulations in \p1 showed that for a decreasing mass-ratio of the
BHs ($q=M_1/M_2 \ge 1$, with $M_1=10^8\,M_\odot=const.$) the fraction
of the ejected stars is increasing. But to extract the bigger amount
of angular momentum from a more massive binary ($L_\BHBidx
=\sqrt{GM_1^3 a}/\sqrt{q(q+1)}$), the increased fraction of ejected
stars is not sufficient, so that the star cluster has to be more
massive in order to absorb the binary's angular momentum and to enable
the BHs to coalesce. Therefore the amount of bound stars has to
increase with the mass $M_2$ of the secondary BH, and so the torus
becomes more massive.

As we have pointed out in the discussion of the influence of the
mass-ratio, the inner regions become increasingly unstable for more
massive secondary BHs, so that at the inner edge of the torus the
stars are moving on almost circular orbits to avoid violent
interactions with the binary. For sufficiently small mass-ratios a
more sharp defined torus forms during the late stages of the merger,
while for young mergers and mergers with large mass-ratios a more
diffuse and shell-like density distribution of the stars is
maintained. This is clearly visible in Figs. 2 and 14 of \p1, which
also show a larger half-opening angle of the torus in the range
$50^\circ$ to $60^\circ$ for $q=1$ (i.e.  $M_2=M_1$) compared to about
$45^\circ$ for $q=10$.  This is due to the stronger torque that is
exerted by the binary with smaller $q$ on stars in the polar cap
region, and therefore less orbits are passing through this region.

In major mergers we also expect the accretion rate on the central BH
to be much higher and thus these mergers to be much more luminous than
minor ones. This is in agreement with the idea that the opening angle
depends on the luminosity, being larger for the more luminous AGN.
Consequently the radiation pressure acting on the torus and its stars
is less in minor mergers, and according to \eqn{eq_est-wrad} the
radius of the stellar winds can extend to much larger radii, not being
stretched into elongated tails. Hence the column density along the LOS
to the center is diminished and the torus becomes optically thin (see
also Sect. \ref{sc_wind_cover}).

This can be illustrated with a simple order-of-magnitude estimate. The
volume of the torus with half-opening angle $\mli{\theta}{trs}$ within
the radial limits $\mli{r}{in}$ and $\mli{r}{out}$ is
$\mli{V}{trs}=\frac{4\pi}{3} (\mli{r}{out}^3
-\mli{r}{in}^3)\cos\mli{\theta}{trs}$. If we distribute the total mass
contained in the stellar winds homogeneously in this volume, the
number density of Hydrogene is
\[
\mli{n}{H} = \frac{\mli{M}{d,
    total}}{1.4\mli{Z}{d}\mli{m}{p}\mli{V}{trs}}.
\]
Integrating along the line of sight from $\mli{r}{in}$ to
$\mli{r}{out}$, yields for the column density of such a homogeneous
torus
\begin{equation}
\label{eq_nhsmooth}
\mli{N}{H} = \frac{3\,\mli{M}{d, total}}{5.6\,\pi
  \mli{Z}{d}\mli{m}{p}\cos\mli{\theta}{trs}}\;
  \frac{\mli{r}{out}-\mli{r}{in}}{\mli{r}{out}^3 - \mli{r}{in}^3}
= 4.8\times 10^{23}\ut{cm}^{-2}\,.
\end{equation}
The total dust mass of the torus is $350\,M_\odot$ (see
\eqn{eq_totaldust}) and for the geometry of the torus we used the same
values as in Sect. \ref{sc_wind_cover}, i.e. $60^\circ$ for the
half-opening angle and $1$ and $5\ut{pc}$ for the inner and outer
radius respectively. This gives less than half the column density of
the patchy torus in Sect. \ref{sc_wind_cover}, when the radiation
pressure is strong enough to turn back the wind into elongated tails.
Thus a torus with smoothly distributed matter would be optically thin,
showing that a sufficiently high luminosity is necessary in order to
maintain an opaque torus.  According to our results from \p1 the
optical thickness of tori as products from major mergers
($M_1=M_2=10^8\,M_\odot$) compared to minor mergers
($M_1=10^8\,M_\odot$, $M_2=10^7$ or $10^6\,M_\odot$) is further
enhanced due to the higher absolute amount of stars that stay bound in
the potential of the binary, even though a larger fraction is ejected.
This is plausible, since in a minor merger a less massive stellar
cluster is surrounding the secondary BH, which is dragged to the
common center by dynamical friction.

Thus we come to the following conclusion: In a major merger with two
BHs of comparable mass ($\sim 10^8\,M_\odot$) more stars are brought
to the common center and stay bound in the potential of the binary,
where they enhance the optical thickness of the torus they constitute.
Because of the stronger torque exerted by the BHs, the half-opening
angle of this torus is larger than in a minor merger. Since the
accretion rate is expected to be close to the Eddington limit, the
stellar winds are exposed to strong radiation pressure from the
center, which shapes the winds into elongated cometary tails, pointing
radially away from the center. Hence they are sufficiently opaque as
to make the stellar torus optically thick. With increasing time the
accretion rate and consequently the luminosity will decrease, so that
the stellar winds in the torus will become less collimated. Therefore
the opacity of the torus will decrease also as an evolutionary effect
of the AGN.

If, on the other hand, a $10^8\,M_\odot$ BH merges with a black hole
of much lower mass, fewer stars are involved and the torus will
contain less stars. Because of the smaller torque of the binary, the
torus is more diffuse and the opening angle is smaller. According to
our simulation, such a torus looks like that of a major merger in
early stages (compare the Figs.~2 and 14 in \p1). In a minor merger
also the accretion rate will be smaller and consequently the
luminosity. This results in less radiation pressure acting on the
stellar winds, which are not shaped into elongated tails, and thus the
opacity of the torus, containing already fewer stars, is even more
diminished.

This seems to be in agreement with the observations: Major mergers of
galaxies of comparable size with BHs in their centers of comparable
mass will violently stir up the stars, gas and dust, and finally
assume an elliptical rather than a spiral shape.  According to our
finding above, ellipticals then should harbour AGN with tori having
larger half-opening angles than spirals and thus on average Type~1
nuclei should be detected more often in elliptical than spiral hosts.
This is supported by the survey of \citet{malkan98}, who find
Seyfert~1 nuclei to reside on average in more early type hosts than
Seyfert~2 nuclei.

It has been emphasized (i.e. \citet{antonucci01b} and references
therein) that there must be a range of covering factors for dusty
tori, with the Type~2-classified objects having higher average
covering factors. It is concluded that the populations are therefore
intrinsically different in their statistical properties to some
extend. Maybe the mass-ratio of the BHs and the evolution of AGN are
the reasons for statistically different covering factors.

If the nucleus happens to be oriented to us in a way that the line of
sight grazes the edge of the torus, the probability to see the nucleus
still directly through a gap in the edge of the clumpy structure is
more probable for large than small mass-ratios. This might be an
explanation for NGC 4151, which is a Type~1 and has an aligned NLR
too. But HST images show a cone on subarcsec scales
\citep{antonucci93}. According to the strict unified model this is not
expected since a Type~1 should only be seen if the observer is inside
the unobscured solid angle so that a projected ionization cone can not
be seen. In our model the patchy torus can have gaps or holes, the
more likely the larger $q$ is and the closer the line of sight comes
to the edge of the torus which is not clearly defined. In the case of
NGC 4151 the observer seems to be outside the opening angle. But
through a gap in the torus close to its edge enough radiation can
escape and is seen directly so that NGC 4151 is classified as a
Type~1.

M87, the nearest giant elliptical galaxy ($z=0.0043$), has been
observed at $10.8\ut{\mu m}$ wavelength by \citet{perlman01}. If there
is a dusty opaque torus it should be visible in the reradiated IR, but
the authors find only little evidence of thermal emission from dust.
Because of its large BH mass ($\sim 3\times 10^9\,M_\odot$,
\citet{marconi97}) and the giant elliptical shape it is likely to have
undergone a major merger in the past. Since then the fuelling of the
AGN probably has decreased and the luminosity dropped to its current
estimated value of $10^{42}\ut{erg/s}$ \citep{whysong01}, what fits to
M87 being on the FR~I-II border, as well in morphology as in radio
power \citep{owen00}.  This luminosity is too weak as to maintain an
opaque torus, which also might have been dissolved since the merger
(see Sect. \ref{sc_lifetime}, lifetime of the torus). A possible minor
merger afterwards most likely would not have sufficiently increased
the luminosity and replenished the torus with stars in order to form a
dusty and opaque torus again.  \citet{whysong01} measured the
reradiation from nuclear hot and warm dust, which must be emitted
according to the unification scheme by the torus that absorbs the
radiation from the AGN it surrounds. This radiation should be found in
almost any hidden AGN and provides an estimate of the unblocked
central luminosity. While for Cyg~A at redshift $0.057$ (more than
$10$ times the redshift of M87), with an estimated luminosity $\sim
1.5\times 10^{45}\ut{erg/s}$, they could detect this component, it is
much weaker in M87, where no hidden nucleus can be found. The
estimated mass accretion in Eddington units
$\dot{M}/\mli{\dot{M}}{edd}$ is in the range $10^{-3.5}$ to $10^{-3}$,
which is much less than $1/50$ \citep{liu99}. Below this value a thin
Shakura-Sunyaev disk can not survive \citep{meyer00}.  With respect to
its large mass, M87 seems to host a starving black hole in the center.

This is in line with the results of \citet{meisenheimer01}, who
compared galaxy-quasar pairs from the 3CR catalogue in the IR range
from $5$ to $180\ut{\mu m}$ to test the unification scheme for
luminous radio galaxies and quasars. The pairs have been selected such
that they match in $178\ut{MHz}$ luminosity, which is thought to be
emitted fairly isotropic, and redshift in order to minimize the
effects of cosmic evolution. They can not distinguish the pairs by
their mid- and far-infrared properties, what strongly supports the
unification scheme. The authors also find the ratio of thermal dust
power $\nu F_\nu$, averaged over $60$ and $100\ut{\mu m}$, to the
radio power at $178\ut{MHz}$ to correlate better with redshift than
with luminosity. They suggest that this might be due to the thermal
power of a radio source being primarily controlled by the accretion
rate, with sources accreting at high rates being more numerous at
large redshifts, when major mergers have been more frequently than
today.

\section{Summary and conclusions}
\label{sc_sum-con}

In \p1 we showed that a central geometrically thick torus, comprised
of stars, results at a distance of order $3\ut{pc}$ from the center as
a product from the merger of two galaxies and their central
supermassive black holes.  This stellar torus has to be about as
massive as the binary black hole in order to enable the BHs to get rid
of their orbital angular momentum and to coalesce on scales of
$10^7\ut{yr}$.

In the present article we proved that this torus with its patchy
structure is in very good agreement with the properties of the
ubiquituous torus in AGN, as are demanded in the unification scheme
and deduced from observations. About 1\% of these stars have strong
enough winds in order to obscure the central radiation source for
lines of sight passing through their winds. The central radiation
pressure shapes the winds of these obscuring stars in the torus into
elongated tails, pointing radially away from the central source.
Consequently their azimuthal extension is about $4\times
10^{-3}\ut{pc}$, such that these winds along their tails are optically
thick, and that their total covering factor within the torus amounts
to $\sim 1$. Thus these winds cause the geometrically thick torus also
to be optically thick with column densities of about
$10^{24}\ut{cm}^{-2}$, just as is observed.  Also the inner radius of
about $1\ut{pc}$ coincides with the evaporation radius of graphite
dust grains \citep{lawrence91}, what has been previously assumed to be
the inner edge, and what is also the distance where the BHs become
hard \citep{milos01} and which defined the inner edge in the
simulation of \p1.

In Sect. \ref{sc_jet-flip} we demonstrated that the recent merging of
two comparable supermassive BHs, the prerequisite for the forming of
the opaque torus, can explain the X-shaped radio galaxies.  According
to the spin-paradigm, fast spinning BHs are powering strong radio jets
in radio galaxies. If such a galaxy merges with another one hosting a
BH of comparable mass, the orbital angular momentum of the resulting
BBH dominates over the maximum possible spin of the primary BH, at
least till their major axis has shrunk to the last stable orbit.  When
the BHs eventually merge, the final spin might be dominated by the
orbital angular momentum, leaving the merged BH spinning close to its
maximum value. As a consequence a new powerful jet emanates into the
direction of the orbital angular momentum, which is aligned with the
symmetry axis of the torus. Therefore this jet streams through the
ambient medium, producing strong radio emission, while the old jet's
lobes, not fed anymore, are slowly fading away.  The time scales of
spectral aging of these lobes is close to the merger-time of the BHs
and thus supports further this interpretation.  The old jet might
intersect with the torus and consequently would interact with its
clumpy medium, also resulting in strong radio emission. This could be
an explanation for the Compact Symmetric Objects, which are thought to
be young and probably evolve into large-size Symmetric Objects, such
as lower luminosity FR~II double radio sources.

With such a patchy torus-model, the BAL~QSOs fit well into the
sequence in which a QSO appears as Blazar if seen pole-on. As the
inclination angle increases it appears as a normal quasar, then high
ionization BAL~QSO, low ionization BAL~QSO and finally, when the line
of sight lies in the equatorial plane of the torus, as ULIRG. For the
intermediate agles the line of sight grazes the surface of the torus,
explaining the features of the BAL~QSOs. In the sample studied by
\citet{gallagher01b}, for most of them a partial-covering absorber
provides a significantly better fit than other models, confirming our
torus model. The detected variability of the BAL quasar PG~2112+059 on
scales of $6\ut{yr}$ is ascribed to changes in the absorber, probably
the column density. This is just the time the stellar winds in the
torus need to move through the line of sight. In Sect.
\ref{sc_bal-trans} we give more examples of objects whose column
densities are observed to change considerably on these time scales.
They are very strongly supporting the idea of the patchy torus, since
the winds of the stars, moving in the potential of the central BH,
have the right size and optical depth at the appropriate distance to
the center to yield such strong variations in $\mli{N}{H}$ on the
right time scales, as they move through the line of sight. Due to the
stellar motion gaps in the torus will be continuously opened and
closed.

For major mergers the resulting torus is more massive and therefore
has a larger column density (see Sect. \ref{sc_massratio}). Because
the accretion-rate and consequently the luminosity will be higher than
in minor mergers, the radiation pressure acting on the winds is
stronger, being able to bend the winds into elongated tails. This
further increases the torus' opacity compared to that of minor
mergers. Since the torque of a more massive binary acting on the stars
is stronger, the opening angle of the resulting torus is wider and
therefore correlates with the central lumnosity, as has been suggested
in the past. But the column density of the torus will also change with
time, as the luminosity decreases when the accretion rate weakens.
Thus the opacity depends on the mass-ratio of the merging black holes
and on evolutionary effects of the AGN. Hence we expect to observe
Type~1 AGN more likely in elliptical host galaxies at larger
redshifts, where major mergers occured more frequently. We also expect
higher covering factors of Type~2 objects. In their survey of Seyfert
galaxies \citet{malkan98} find on average Type~1 AGN to be hosted by
more early type galaxies than Seyfert~2s, in agreement with our
reasoning.

Previous compact torus models with a smooth distribution of the matter
all faced the same problem in predicting a too narrow infrared
spectrum, and to fit the data, additional NIR sources had to be
invoked \citep{pier93,granato94,efstathiou95,alonso-herrero01}. As has
been already mentioned in these papers, it is obvious that a patchy
torus will tend to increase the dust temperature in the outer parts,
since they are exposed to the radiation from the center shining
through gaps in the inner parts of the torus. This will tend to
broaden the IR spectrum. Such models also exhibit strong emission or
absorption $9.7\ut{\mu m}$ Silicate features, which are usually not in
agreement with the observations. Another problem is to achieve the
geometrical thickness of the torus, which according to \citet{pier92b}
can be supported by radiation pressure. But for a low luminosity
source they need the torus to be clumpy and introduce a radiation
pressure driven random motion of the clumps in order to maintain the
thickness of the torus.

In the model we proposed here and in \p1, the torus is naturally thick
as a result of the stars moving in the potential of the binary.  The
patchy structure inevitably leads to self-shielding of the stellar
winds and also within the winds. This probably has an effect on the
temperature of the gas and dust as a function of the distance to the
central source. Both, the selfshielding and the temperature
distribution in the torus will have a strong influence on the
reprocessed and reemitted radiation. The calculation of the
temperature distribution and spectra of the winds and the torus they
constitute would involve a full threedimensional treatment, so that
radiation-transfer calculations are well beyond the scope of this
paper and we can not give a comment on possible Silicate features as
function of the inclination angle.

The torus model proposed in these two articles gives a coherent
picture with respect to the formation of the torus, its evolution and
application to observations in terms of the unification scheme.

\begin{acknowledgements}
  This article as well as \p1 are based on on the first author's
  Ph.D.~thesis from june 2000 \citep{zier00}.

  CZ would like to thank R.R.J.~Antonucci, M.~Malkan, D.~Merritt,
  G.~Smith, and S.~Westerhoff for their helpful discussions and their
  generous and kind hospitality (fall 2000). CZ also acknowledges the
  longterm support by the MPIfR. PLB would like to thank Drs.~A.~Donea
  and R.~Protheroe for extended discussion on tori, and their
  hospitality at Adelaide. PLB also acknowledges the discussions with
  R.R.J.~Antonucci, M.~Malkan, G.~Sch{\"a}fer and N.~Straumann. PLB
  and CZ also like to especially thank M.~Chirvasa for the discussions
  on her work of gravitational radiation of a black hole binary, her
  Ph.D.~thesis at the University of Bukarest of february 2000
  \citep{chirvasa02}. High energy physics in PLB's group is supported
  by AUGER-Theory grant 05~CU1ERA/3 from DESY/BMBF.

\end{acknowledgements}

\begin{appendix}
\end{appendix}

\bibliography{refs}

\begin{thebibliography}{117}
\expandafter\ifx\csname natexlab\endcsname\relax\def\natexlab#1{#1}\fi

\bibitem[{{Alexander} \& {Netzer}(1994)}]{alexander_netzer94}
{Alexander}, T. \& {Netzer}, H. 1994, \mnras, 270, 781

\bibitem[{{Alexander} \& {Netzer}(1997)}]{alexander_netzer97}
---. 1997, \mnras, 284, 967

\bibitem[{{Alonso-Herrero} {et~al.}(2001){Alonso-Herrero}, {Quillen},
  {Simpson}, {Efstathiou}, \& {Ward}}]{alonso-herrero01}
{Alonso-Herrero}, A., {Quillen}, A.~C., {Simpson}, C., {Efstathiou}, A., \&
  {Ward}, M.~J. 2001, \aj, 121, 1369

\bibitem[{{Antonucci}(1993)}]{antonucci93}
{Antonucci}, R. 1993, \araa, 31, 473

\bibitem[{{Antonucci}(2001{\natexlab{a}})}]{antonucci01b}
{Antonucci}, R.~R.~J. 2001{\natexlab{a}}, \texttt{astro-ph/0110343}

\bibitem[{{Antonucci}(2001{\natexlab{b}})}]{antonucci01}
---. 2001{\natexlab{b}}, \texttt{astro-ph/0103048}

\bibitem[{{Antonucci} \& {Miller}(1985)}]{antonucci85}
{Antonucci}, R. R.~J. \& {Miller}, J.~S. 1985, \apj, 297, 621

\bibitem[{{Baribaud} {et~al.}(1992){Baribaud}, {Alloin}, {Glass}, \&
  {Pelat}}]{baribaud92}
{Baribaud}, T., {Alloin}, D., {Glass}, I., \& {Pelat}, D. 1992, \aap, 256, 375

\bibitem[{{Barthel}(1989)}]{barthel89}
{Barthel}, P.~D. 1989, \apj, 336, 606

\bibitem[{{Barvainis}(1987)}]{barvainis87}
{Barvainis}, R. 1987, \apj, 320, 537

\bibitem[{{Biermann} \& {Harwit}(1980)}]{biermann80}
{Biermann}, P. \& {Harwit}, M. 1980, \apjl, 241, L105

\bibitem[{{Biermann} \& {Shapiro}(1979)}]{biermann79}
{Biermann}, P. \& {Shapiro}, S.~L. 1979, \apjl, 230, L33

\bibitem[{{Biermann}(1989)}]{biermann89}
{Biermann}, P.~L. 1989, in NATO ASI series. Ser. C, Mathematical and physical
  sciences; vol. 270: Cosmic Gamma Rays, Neutrinos, and Related Astrophysics,
  21+

\bibitem[{{Boller} {et~al.}(2002){Boller}, {Fabian}, {Sunyaev}, {Tr{\" u}mper},
  {Vaughan}, {Ballantyne}, {Brandt}, {Keil}, \& {Iwasawa}}]{boller02}
{Boller}, T., {Fabian}, A.~C., {Sunyaev}, R., {et~al.} 2002, \mnras, 329, L1

\bibitem[{{Brotherton} {et~al.}(1997){Brotherton}, {Tran}, {van Breugel},
  {Dey}, \& {Antonucci}}]{brotherton97}
{Brotherton}, M.~S., {Tran}, H.~D., {van Breugel}, W., {Dey}, A., \&
  {Antonucci}, R. 1997, \apjl, 487, L113

\bibitem[{{Chini} {et~al.}(1989){Chini}, {Kreysa}, \& {Biermann}}]{chini89}
{Chini}, R., {Kreysa}, E., \& {Biermann}, P.~L. 1989, \aap, 219, 87

\bibitem[{{Chirvasa}(2002)}]{chirvasa02}
{Chirvasa}, M. 2002, M.Sc.~Thesis, University of Bukarest

\bibitem[{{Clavel} {et~al.}(1989){Clavel}, {Wamsteker}, \& {Glass}}]{clavel89}
{Clavel}, J., {Wamsteker}, W., \& {Glass}, I.~S. 1989, \apj, 337, 236

\bibitem[{{Conway} \& {Blanco}(1995)}]{conway95}
{Conway}, J.~E. \& {Blanco}, P.~R. 1995, \apjl, 449, L131

\bibitem[{{Edwards}(1980)}]{edwards80}
{Edwards}, A.~C. 1980, \mnras, 190, 757

\bibitem[{{Efstathiou} \& {Rowan-Robinson}(1995)}]{efstathiou95}
{Efstathiou}, A. \& {Rowan-Robinson}, M. 1995, \mnras, 273, 649

\bibitem[{{Egami} {et~al.}(1996){Egami}, {Iwamuro}, {Maihara}, {Oya}, \&
  {Cowie}}]{egami96}
{Egami}, E., {Iwamuro}, F., {Maihara}, T., {Oya}, S., \& {Cowie}, L.~L. 1996,
  \aj, 112, 73+

\bibitem[{{Fanti} {et~al.}(1995){Fanti}, {Fanti}, {Dallacasa}, {Schilizzi},
  {Spencer}, \& {Stanghellini}}]{fanti95}
{Fanti}, C., {Fanti}, R., {Dallacasa}, D., {et~al.} 1995, \aap, 302, 317+

\bibitem[{{Farouki} \& {Shapiro}(1982)}]{farouki82}
{Farouki}, R.~T. \& {Shapiro}, S.~L. 1982, \apj, 259, 103

\bibitem[{{Gallagher} {et~al.}(2002){Gallagher}, {Brandt}, {Chartas}, \&
  {Garmire}}]{gallagher01b}
{Gallagher}, S.~C., {Brandt}, W.~N., {Chartas}, G., \& {Garmire}, G.~P. 2002,
  \apj, 567, 37

\bibitem[{{Gallagher} {et~al.}(2001){Gallagher}, {Brandt}, {Laor}, {Elvis},
  {Mathur}, {Wills}, \& {Iyomoto}}]{gallagher01a}
{Gallagher}, S.~C., {Brandt}, W.~N., {Laor}, A., {et~al.} 2001, \apj, 546, 795

\bibitem[{{George} {et~al.}(1998){George}, {Mushotzky}, {Turner}, {Yaqoob},
  {Ptak}, {Nandra}, \& {Netzer}}]{george98}
{George}, I.~M., {Mushotzky}, R., {Turner}, T.~J., {et~al.} 1998, \apj, 509,
  146

\bibitem[{{Gilli} {et~al.}(2000){Gilli}, {Maiolino}, {Marconi}, {Risaliti},
  {Dadina}, {Weaver}, \& {Colbert}}]{gilli00}
{Gilli}, R., {Maiolino}, R., {Marconi}, A., {et~al.} 2000, \aap, 355, 485

\bibitem[{{Granato} \& {Danese}(1994)}]{granato94}
{Granato}, G.~L. \& {Danese}, L. 1994, \mnras, 268, 235+

\bibitem[{{Green} {et~al.}(2001){Green}, {Aldcroft}, {Mathur}, {Wilkes}, \&
  {Elvis}}]{green01}
{Green}, P.~J., {Aldcroft}, T.~L., {Mathur}, S., {Wilkes}, B.~J., \& {Elvis},
  M. 2001, \apj, 558, 109

\bibitem[{{Gregg} {et~al.}(2000){Gregg}, {Becker}, {Brotherton},
  {Laurent-Muehleisen}, {Lacy}, \& {White}}]{gregg00}
{Gregg}, M.~D., {Becker}, R.~H., {Brotherton}, M.~S., {et~al.} 2000, \apj, 544,
  142

\bibitem[{{Guainazzi} {et~al.}(1998){Guainazzi}, {Nicastro}, {Fiore}, {Matt},
  {McHardy}, {Orr}, {Barr}, {Fruscione}, {Papadakis}, {Parmar}, {Uttley},
  {Perola}, \& {Piro}}]{guainazzi98}
{Guainazzi}, M., {Nicastro}, F., {Fiore}, F., {et~al.} 1998, \mnras, 301, L1

\bibitem[{{Haas} {et~al.}(2000){Haas}, {M{\"u}ller}, {Chini}, {Meisenheimer},
  {Klaas}, {Lemke}, {Kreysa}, \& {Camenzind}}]{haas00}
{Haas}, M., {M{\"u}ller}, S. A.~H., {Chini}, R., {et~al.} 2000, \aap, 354, 453

\bibitem[{{Halpern}(1982)}]{halpern82}
{Halpern}, J.~P. 1982, Ph.D.~Thesis, 4+

\bibitem[{{Hoefner} {et~al.}(1996){Hoefner}, {Fleischer}, {Gauger},
  {Feuchtinger}, {Dorfi}, {Winters}, \& {Sedlmayr}}]{hoefner96}
{Hoefner}, S., {Fleischer}, A.~J., {Gauger}, A., {et~al.} 1996, \aap, 314, 204

\bibitem[{{Huchra} \& {Burg}(1992)}]{huchra92}
{Huchra}, J. \& {Burg}, R. 1992, \apj, 393, 90

\bibitem[{{Kinney} {et~al.}(1991){Kinney}, {Antonucci}, {Ward}, {Wilson}, \&
  {Whittle}}]{kinney91}
{Kinney}, A.~L., {Antonucci}, R.~R.~J., {Ward}, M.~J., {Wilson}, A.~S., \&
  {Whittle}, M. 1991, \apj, 377, 100

\bibitem[{{Knapp} \& {Morris}(1985)}]{knapp85}
{Knapp}, G.~R. \& {Morris}, M. 1985, \apj, 292, 640

\bibitem[{{Krichbaum} {et~al.}(1998){Krichbaum}, {Alef}, {Witzel}, {Zensus},
  {Booth}, {Greve}, \& {Rogers}}]{krichbaum98}
{Krichbaum}, T.~P., {Alef}, W., {Witzel}, A., {et~al.} 1998, \aap, 329, 873

\bibitem[{{Krolik} \& {Begelman}(1988)}]{krolik88}
{Krolik}, J.~H. \& {Begelman}, M.~C. 1988, \apj, 329, 702

\bibitem[{{Lawrence}(1991)}]{lawrence91}
{Lawrence}, A. 1991, \mnras, 252, 586

\bibitem[{{Lawrence} \& {Elvis}(1982)}]{lawrence82}
{Lawrence}, A. \& {Elvis}, M. 1982, \apj, 256, 410

\bibitem[{{Liu} {et~al.}(1999){Liu}, {Yuan}, {Meyer}, {Meyer-Hofmeister}, \&
  {Xie}}]{liu99}
{Liu}, B.~F., {Yuan}, W., {Meyer}, F., {Meyer-Hofmeister}, E., \& {Xie}, G.~Z.
  1999, \apjl, 527, L17

\bibitem[{{Longair}(1981)}]{longair81}
{Longair}, M.~S. 1981, {High energy astrophysics} (Cambridge: University Press,
  1981)

\bibitem[{{Lovelace} {et~al.}(1998){Lovelace}, {Romanova}, \&
  {Biermann}}]{lovelace98}
{Lovelace}, R.~V.~E., {Romanova}, M.~M., \& {Biermann}, P.~L. 1998, \aap, 338,
  856

\bibitem[{{Macchetto} {et~al.}(1994){Macchetto}, {Capetti}, {Sparks}, {Axon},
  \& {Boksenberg}}]{macchetto94}
{Macchetto}, F., {Capetti}, A., {Sparks}, W.~B., {Axon}, D.~J., \&
  {Boksenberg}, A. 1994, \apjl, 435, L15

\bibitem[{{MacDonald} {et~al.}(1991){MacDonald}, {Stanev}, \&
  {Biermann}}]{macdonald91}
{MacDonald}, J., {Stanev}, T., \& {Biermann}, P.~L. 1991, \apj, 378, 30

\bibitem[{{Maiolino} {et~al.}(1998){Maiolino}, {Salvati}, {Bassani}, {Dadina},
  {della Ceca}, {Matt}, {Risaliti}, \& {Zamorani}}]{maiolino98}
{Maiolino}, R., {Salvati}, M., {Bassani}, L., {et~al.} 1998, \aap, 338, 781

\bibitem[{{Malkan} {et~al.}(1998){Malkan}, {Gorjian}, \& {Tam}}]{malkan98}
{Malkan}, M.~A., {Gorjian}, V., \& {Tam}, R. 1998, \apjs, 117, 25+

\bibitem[{{Marconi} {et~al.}(1997){Marconi}, {Axon}, {Macchetto}, {Capetti},
  {Soarks}, \& {Crane}}]{marconi97}
{Marconi}, A., {Axon}, D.~J., {Macchetto}, F.~D., {et~al.} 1997, \mnras, 289,
  L21

\bibitem[{{Mathis} {et~al.}(1977){Mathis}, {Rumpl}, \& {Nordsieck}}]{mathis77}
{Mathis}, J.~S., {Rumpl}, W., \& {Nordsieck}, K.~H. 1977, \apj, 217, 425

\bibitem[{{Meier}(2001)}]{meier01}
{Meier}, D.~L. 2001, \apjl, 548, L9

\bibitem[{{Meisenheimer} {et~al.}(2001){Meisenheimer}, {Haas}, {M{\" u}ller},
  {Chini}, {Klaas}, \& {Lemke}}]{meisenheimer01}
{Meisenheimer}, K., {Haas}, M., {M{\" u}ller}, S.~A.~H., {et~al.} 2001, \aap,
  372, 719

\bibitem[{{Merritt} \& {Ferrarese}(2001{\natexlab{a}})}]{merritt01a}
{Merritt}, D. \& {Ferrarese}, L. 2001{\natexlab{a}}, \mnras, 320, L30

\bibitem[{{Merritt} \& {Ferrarese}(2001{\natexlab{b}})}]{merritt01b}
---. 2001{\natexlab{b}}, \apj, 547, 140

\bibitem[{{Meyer} {et~al.}(2000){Meyer}, {Liu}, \&
  {Meyer-Hofmeister}}]{meyer00}
{Meyer}, F., {Liu}, B.~F., \& {Meyer-Hofmeister}, E. 2000, \aap, 354, L67

\bibitem[{{Miller} {et~al.}(1991){Miller}, {Goodrich}, \& {Mathews}}]{miller91}
{Miller}, J.~S., {Goodrich}, R.~W., \& {Mathews}, W.~G. 1991, \apj, 378, 47

\bibitem[{{Milosavljevi{\' c}} \& {Merritt}(2001)}]{milos01}
{Milosavljevi{\' c}}, M.~. \& {Merritt}, D. 2001, \apj, 563, 34

\bibitem[{{Mulchaey} {et~al.}(1992){Mulchaey}, {Mushotzky}, \&
  {Weaver}}]{mulchaey92}
{Mulchaey}, J.~S., {Mushotzky}, R.~F., \& {Weaver}, K.~A. 1992, \apjl, 390, L69

\bibitem[{{Mushotzky}(1982)}]{mushotzky82}
{Mushotzky}, R.~F. 1982, \apj, 256, 92

\bibitem[{{Nagar} \& {Wilson}(1999)}]{nagar99}
{Nagar}, N.~M. \& {Wilson}, A.~S. 1999, \apj, 516, 97

\bibitem[{{Nandra} \& {Pounds}(1994)}]{nandra94}
{Nandra}, K. \& {Pounds}, K.~A. 1994, \mnras, 268, 405+

\bibitem[{{Niemeyer} \& {Biermann}(1993)}]{niemeyer93}
{Niemeyer}, M. \& {Biermann}, P.~L. 1993, \aap, 279, 393

\bibitem[{{Oliva} {et~al.}(1995){Oliva}, {Origlia}, {Kotilainen}, \&
  {Moorwood}}]{oliva95}
{Oliva}, E., {Origlia}, L., {Kotilainen}, J.~K., \& {Moorwood}, A. F.~M. 1995,
  \aap, 301, 55+

\bibitem[{{Osterbrock} \& {Shaw}(1988)}]{osterbrock88}
{Osterbrock}, D.~E. \& {Shaw}, R.~A. 1988, \apj, 327, 89

\bibitem[{{Owen} {et~al.}(2000){Owen}, {Eilek}, \& {Kassim}}]{owen00}
{Owen}, F.~N., {Eilek}, J.~A., \& {Kassim}, N.~E. 2000, \apj, 543, 611

\bibitem[{{Owsianik} \& {Conway}(1998)}]{owsianik98b}
{Owsianik}, I. \& {Conway}, J.~E. 1998, \aap, 337, 69

\bibitem[{{Owsianik} {et~al.}(1998){Owsianik}, {Conway}, \&
  {Polatidis}}]{owsianik98a}
{Owsianik}, I., {Conway}, J.~E., \& {Polatidis}, A.~G. 1998, \aap, 336, L37

\bibitem[{{Owsianik} {et~al.}(1999){Owsianik}, {Conway}, \&
  {Polatidis}}]{owsianik99}
---. 1999, New Astronomy Review, 43, 669

\bibitem[{{P{\' e}rez} {et~al.}(2000){P{\' e}rez}, {M{\' a}rquez}, {Marrero},
  {Durret}, {Gonz{\' a}lez Delgado}, {Masegosa}, {Maza}, \& {Moles}}]{perez00}
{P{\' e}rez}, E., {M{\' a}rquez}, I., {Marrero}, I., {et~al.} 2000, \aap, 353,
  893

\bibitem[{{Parma} {et~al.}(1985){Parma}, {Ekers}, \& {Fanti}}]{parma85}
{Parma}, P., {Ekers}, R.~D., \& {Fanti}, R. 1985, \aaps, 59, 511

\bibitem[{{Perlman} {et~al.}(2001){Perlman}, {Sparks}, {Radomski}, {Packham},
  {Fisher}, {Pi{\~ n}a}, \& {Biretta}}]{perlman01}
{Perlman}, E.~S., {Sparks}, W.~B., {Radomski}, J., {et~al.} 2001, \apjl, 561,
  L51

\bibitem[{{Peterson}(1993)}]{peterson93}
{Peterson}, B.~M. 1993, \pasp, 105, 247

\bibitem[{{Phillips} \& {Mutel}(1982)}]{phillips82}
{Phillips}, R.~B. \& {Mutel}, R.~L. 1982, \aap, 106, 21

\bibitem[{{Pier} \& {Krolik}(1992{\natexlab{a}})}]{pier92}
{Pier}, E.~A. \& {Krolik}, J.~H. 1992{\natexlab{a}}, \apj, 401, 99

\bibitem[{{Pier} \& {Krolik}(1992{\natexlab{b}})}]{pier92b}
---. 1992{\natexlab{b}}, \apjl, 399, L23

\bibitem[{{Pier} \& {Krolik}(1993)}]{pier93}
---. 1993, \apj, 418, 673+

\bibitem[{{Pogge}(1989)}]{pogge89}
{Pogge}, R.~W. 1989, \apj, 345, 730

\bibitem[{{Readhead} {et~al.}(1996){Readhead}, {Taylor}, {Pearson}, \&
  {Wilkinson}}]{readhead96}
{Readhead}, A.~C.~S., {Taylor}, G.~B., {Pearson}, T.~J., \& {Wilkinson}, P.~N.
  1996, \apj, 460, 634+

\bibitem[{{Risaliti} {et~al.}(2000){Risaliti}, {Maiolino}, \&
  {Bassani}}]{risaliti00}
{Risaliti}, G., {Maiolino}, R., \& {Bassani}, L. 2000, \aap, 356, 33

\bibitem[{{Rottmann}(2001)}]{rottmann01}
{Rottmann}, H. 2001, Ph.D.~Thesis, University of Bonn

\bibitem[{{Rottmann} {et~al.}(1998){Rottmann}, {Dennet-Thorpe}, \&
  {Klein}}]{rottmann98}
{Rottmann}, H., {Dennet-Thorpe}, J., \& {Klein}, U. 1998, Astronomische
  Gesellschaft Meeting Abstracts, Abstracts of Contributed Talks and Posters
  presented at the Annual Scientific Meeting of the Astronomische Gesellschaft
  at Heidelberg, September 14--19, 1998, poster \#P77, 14, 77+

\bibitem[{{Rybicki} \& {Lightman}(1979)}]{rybicki79}
{Rybicki}, G.~B. \& {Lightman}, A.~P. 1979, {Radiative Processes in
  Astrophysics} (John Wiley \& Sons)

\bibitem[{{Sanders} {et~al.}(1989){Sanders}, {Phinney}, {Neugebauer}, {Soifer},
  \& {Matthews}}]{sanders89b}
{Sanders}, D.~B., {Phinney}, E.~S., {Neugebauer}, G., {Soifer}, B.~T., \&
  {Matthews}, K. 1989, \apj, 347, 29

\bibitem[{{Schmidt} \& {Hines}(1999)}]{schmidt99}
{Schmidt}, G.~D. \& {Hines}, D.~C. 1999, \apj, 512, 125

\bibitem[{{Shull}(1983)}]{shull83}
{Shull}, J.~M. 1983, \apj, 264, 446

\bibitem[{{Singh} {et~al.}(1985){Singh}, {Garmire}, \& {Nousek}}]{singh85}
{Singh}, K.~P., {Garmire}, G.~P., \& {Nousek}, J. 1985, \apj, 297, 633

\bibitem[{{Sitko}(1991)}]{sitko91}
{Sitko}, M. 1991, in Variability of Active Galactic Nuclei, ed. H.~R. {Miller}
  \& P.~J. {Wiita} (Cambridge University Press), 104

\bibitem[{{Sprayberry} \& {Foltz}(1992)}]{sprayberry92}
{Sprayberry}, D. \& {Foltz}, C.~B. 1992, \apj, 390, 39

\bibitem[{{Stecker} {et~al.}(1991){Stecker}, {Done}, {Salamon}, \&
  {Sommers}}]{stecker91}
{Stecker}, F.~W., {Done}, C., {Salamon}, M.~H., \& {Sommers}, P. 1991, Physical
  Review Letters, 66, 2697

\bibitem[{{Taniguchi} \& {Anabuki}(1999)}]{taniguchi99a}
{Taniguchi}, Y. \& {Anabuki}, N. 1999, \apjl, 521, L103

\bibitem[{{Tinsley} \& {Larson}(1977)}]{toomre77}
{Tinsley}, B.~M. \& {Larson}, R.~B., eds. 1977, {The Evolution of galaxies and
  stellar populations}, ed. B.~M. {Tinsley} \& R.~B. {Larson}

\bibitem[{{Tout} {et~al.}(1989){Tout}, {Eggleton}, {Fabian}, \&
  {Pringle}}]{tout89}
{Tout}, C.~A., {Eggleton}, P.~P., {Fabian}, A.~C., \& {Pringle}, J.~E. 1989,
  \mnras, 238, 427

\bibitem[{{Turner} {et~al.}(2000){Turner}, {Perola}, {Fiore}, {Matt}, {George},
  {Piro}, \& {Bassani}}]{turner00}
{Turner}, T.~J., {Perola}, G.~C., {Fiore}, F., {et~al.} 2000, \apj, 531, 245

\bibitem[{{Turner} \& {Pounds}(1989)}]{turner89}
{Turner}, T.~J. \& {Pounds}, K.~A. 1989, \mnras, 240, 833

\bibitem[{{Turner} {et~al.}(1991){Turner}, {Weaver}, {Mushotzky}, {Holt}, \&
  {Madejski}}]{turner91}
{Turner}, T.~J., {Weaver}, K.~A., {Mushotzky}, R.~F., {Holt}, S.~S., \&
  {Madejski}, G.~M. 1991, \apj, 381, 85

\bibitem[{{Ueno} {et~al.}(1994){Ueno}, {Koyama}, {Nishida}, {Yamauchi}, \&
  {Ward}}]{ueno94}
{Ueno}, S., {Koyama}, K., {Nishida}, M., {Yamauchi}, S., \& {Ward}, M.~J. 1994,
  \apjl, 431, L1

\bibitem[{{Urry} \& {Padovani}(1995)}]{urry95}
{Urry}, C.~M. \& {Padovani}, P. 1995, \pasp, 107, 803+

\bibitem[{{Uttley} {et~al.}(1998){Uttley}, {McHardy}, {Papadakis}, {Cagnoni},
  \& {Fruscione}}]{uttley98}
{Uttley}, P., {McHardy}, I.~M., {Papadakis}, I.~E., {Cagnoni}, I., \&
  {Fruscione}, A. 1998, in The Active X-ray Sky: Results from BeppoSAX and
  RXTE. Proceedings of the Active X-ray Sky symposium, October 21-24, 1997,
  Rome, Italy, Edited by L. Scarsi, H. Bradt, P. Giommi, and F. Fiore.
  Publisher: Amsterdam: Elsevier, 1998. Reprinted from: Nuclear Physics B,
  (Proc. Suppl.), vol. 69/1-3. ISBN: 0444829903., p.490, 490+

\bibitem[{{Voit} \& {Shull}(1988)}]{voit88}
{Voit}, G.~M. \& {Shull}, J.~M. 1988, \apj, 331, 197

\bibitem[{{Voit} {et~al.}(1993){Voit}, {Weymann}, \& {Korista}}]{voit93}
{Voit}, G.~M., {Weymann}, R.~J., \& {Korista}, K.~T. 1993, \apj, 413, 95

\bibitem[{{Warwick} {et~al.}(1993){Warwick}, {Sembay}, {Yaqoob}, {Makishima},
  {Ohashi}, {Tashiro}, \& {Kohmura}}]{warwick93}
{Warwick}, R.~S., {Sembay}, S., {Yaqoob}, T., {et~al.} 1993, \mnras, 265, 412+

\bibitem[{{Weaver} {et~al.}(1996){Weaver}, {Nousek}, {Yaqoob}, {Mushotzky},
  {Makino}, \& {Otani}}]{weaver96}
{Weaver}, K.~A., {Nousek}, J., {Yaqoob}, T., {et~al.} 1996, \apj, 458, 160+

\bibitem[{{Weymann}(1997)}]{weymann97}
{Weymann}, R. 1997, in ASP Conf. Ser. 128: Mass Ejection from Active Galactic
  Nuclei, 3+

\bibitem[{{Weymann} {et~al.}(1991){Weymann}, {Morris}, {Foltz}, \&
  {Hewett}}]{weymann91}
{Weymann}, R.~J., {Morris}, S.~L., {Foltz}, C.~B., \& {Hewett}, P.~C. 1991,
  \apj, 373, 23

\bibitem[{{Whysong} \& {Antonucci}(2001)}]{whysong01}
{Whysong}, D. \& {Antonucci}, R.~R.~J. 2001, \texttt{astro-ph/0106381}

\bibitem[{{Wilkinson} {et~al.}(1994){Wilkinson}, {Polatidis}, {Readhead}, {Xu},
  \& {Pearson}}]{wilkinson94}
{Wilkinson}, P.~N., {Polatidis}, A.~G., {Readhead}, A.~C.~S., {Xu}, W., \&
  {Pearson}, T.~J. 1994, \apjl, 432, L87

\bibitem[{{Willott} {et~al.}(1999){Willott}, {Rawlings}, \&
  {Blundell}}]{willott99}
{Willott}, C.~J., {Rawlings}, S., \& {Blundell}, K.~M. 1999, in ASP Conf. Ser.
  162: Quasars and Cosmology, 135+

\bibitem[{{Wills}(1999)}]{wills99}
{Wills}, B.~J. 1999, in ASP Conf. Ser. 162: Quasars and Cosmology, 101+

\bibitem[{{Wilson} {et~al.}(1993){Wilson}, {Braatz}, {Heckman}, {Krolik}, \&
  {Miley}}]{wilson93}
{Wilson}, A.~S., {Braatz}, J.~A., {Heckman}, T.~M., {Krolik}, J.~H., \&
  {Miley}, G.~K. 1993, \apjl, 419, L61

\bibitem[{{Wilson} \& {Colbert}(1995)}]{wilson95}
{Wilson}, A.~S. \& {Colbert}, E. J.~M. 1995, \apj, 438, 62

\bibitem[{{Wilson} \& {Tsvetanov}(1994)}]{wilson94}
{Wilson}, A.~S. \& {Tsvetanov}, Z.~I. 1994, \aj, 107, 1227

\bibitem[{{Winters} {et~al.}(1994){Winters}, {Dominik}, \&
  {Sedlmayr}}]{winters94}
{Winters}, J.~M., {Dominik}, C., \& {Sedlmayr}, E. 1994, \aap, 288, 255

\bibitem[{{Xue} {et~al.}(1998){Xue}, {Otani}, {Mihara}, {Cappi}, \&
  {Matsuoka}}]{xue98}
{Xue}, S., {Otani}, C., {Mihara}, T., {Cappi}, M., \& {Matsuoka}, M. 1998,
  \pasj, 50, 519

\bibitem[{{Young} {et~al.}(1977){Young}, {Shields}, \& {Wheeler}}]{young77}
{Young}, P.~J., {Shields}, G.~A., \& {Wheeler}, J.~C. 1977, \apj, 212, 367

\bibitem[{{Zier}(2000)}]{zier00}
{Zier}, C. 2000, Ph.D.~Thesis, University of Bonn

\bibitem[{{Zier} \& {Biermann}(2001)}]{zier01}
{Zier}, C. \& {Biermann}, P.~L. 2001, \aap, 377, 23, {(\p1)}

\end{thebibliography}
\bibliographystyle{aa}

\end{document}